\documentclass[12pt]{article}
\usepackage{epsfig,amsmath,euscript,graphicx,psfrag}

\newcommand{\gw}{0.15\textwidth}
\newcommand{\ph}{0.06\textwidth}

\def\delslash{\rlap{\hspace{0.02cm}/}{\partial}}
\def\nslash{\rlap{\hspace{0.02cm}/}{n}}
\def\nbslash{\rlap{\hspace{0.02cm}/}{\bar n}}
\def\xslash{\rlap{\hspace{0.02cm}/}{x}}
\def\yslash{\rlap{\hspace{0.02cm}/}{y}}
\def\zslash{\rlap{\hspace{0.02cm}/}{z}}
\def\Dslash{\rlap{\hspace{0.07cm}/}{D}}
\def\Aslash{\rlap{\hspace{0.07cm}/}{A}}
\def\calAslash{\rlap{\hspace{0.08cm}/}{{\EuScript A}}}
\def\calDslash{\rlap{\hspace{0.1cm}/}{{\EuScript D}}}

\def\A{{\EuScript A}}
\def\D{{\EuScript D}}
\def\Q{{\EuScript Q}}
\def\X{{\EuScript X}}

\begin{document}

\begin{titlepage}

\begin{flushright}
CLNS~03/1836\\
SLAC--PUB--10098\\
{\tt hep-ph/0308122}\\[0.2cm]
\end{flushright}

\vspace{0.7cm}
\begin{center}
\Large\bf 
Soft-Collinear Messengers:
A New Mode in Soft-Collinear Effective Theory
\end{center}

\vspace{0.8cm}
\begin{center}
{\sc Thomas Becher$^{(a)}$, Richard J.~Hill$^{(a)}$ and Matthias 
Neubert$^{(b)}$}\\
\vspace{0.7cm}
$^{(a)}${\sl Stanford Linear Accelerator Center, Stanford University\\
Stanford, CA 94309, U.S.A.} \\
\vspace{0.3cm}
$^{(b)}${\sl Newman Laboratory for Elementary-Particle Physics, 
Cornell University\\
Ithaca, NY 14853, U.S.A.}
\end{center}

\vspace{1.0cm}
\begin{abstract}
\vspace{0.2cm}\noindent 
It is argued that soft-collinear effective theory for processes involving 
both soft and collinear partons, such as exclusive $B$-meson decays, 
should include a new mode in addition to soft and collinear ones. These 
``soft-collinear messengers'' can interact with both soft and collinear 
particles without taking them far off-shell. They thus can communicate 
between the soft and collinear sectors of the theory. The relevance of 
the new mode is demonstrated with an explicit example, and the formalism 
incorporating the corresponding quark and gluon fields into the effective 
Lagrangian is developed.
\end{abstract}
\vfil

\end{titlepage}

\section{Introduction}

There is currently much effort devoted to applications of soft-collinear 
effective theory (SCET) 
\cite{Bauer:2000yr,Bauer:2001yt,Chay:2002vy,Beneke:2002ph,Hill:2002vw} to 
exclusive $B$ decays
\cite{Bauer:2001cu,Lunghi:2002ju,Bauer:2002aj,Bosch:2003fc,Chay:2003ju}. 
SCET provides a systematic framework in which to discuss QCD 
factorization theorems for these processes \cite{Beneke:1999br} and 
power corrections using the language of effective field theory. The 
hadronic states relevant to exclusive decays such as $B\to K^*\gamma$ or 
$B\to\pi\pi$ contain highly energetic, collinear partons inside the 
final-state light mesons, and soft partons inside the initial $B$ meson. 
Understanding the intricate interplay between soft and collinear degrees 
of freedom is a challenge that one hopes to address using the effective 
theory. This interplay is relevant even in simpler processes such as 
semileptonic $B$ decays near $q^2\approx 0$, which are described in terms 
of heavy-to-light form factors at large recoil.

Power counting in SCET is based on an expansion parameter 
$\lambda\sim\Lambda/E$, where $E\gg\Lambda_{\rm QCD}$ is a large scale
(typically $E\sim m_b$ in $B$ decays), and $\Lambda\sim\Lambda_{\rm QCD}$ 
is of order the QCD scale. A complication in SCET is that different 
components of particle momenta and fields may scale differently with the 
large scale $E$. To make this scaling explicit one introduces two
light-like vectors $n^\mu$ and $\bar n^\mu$ satisfying $n^2=\bar n^2=0$
and $n\cdot\bar n=2$. Typically, $n^\mu$ is the direction of an outgoing
fast hadron (or a jet of hadrons). Any 4-vector can then be decomposed as
\begin{equation}
   p^\mu = (n\cdot p)\,\frac{\bar n^\mu}{2}
    + (\bar n\cdot p)\,\frac{n^\mu}{2} + p_\perp^\mu
   \equiv p_+^\mu + p_-^\mu + p_\perp^\mu \,,
\end{equation}
where $p_\perp\cdot n=p_\perp\cdot\bar n=0$. The light-like vectors
$p_\pm^\mu$ are defined by this relation. The relevant SCET degrees of 
freedom describing the partons in the external hadronic states of 
exclusive $B$ decays are soft and collinear, where 
$p_s^\mu\sim E(\lambda,\lambda,\lambda)$ for soft momenta and 
$p_c^\mu\sim E(\lambda^2,1,\lambda)$ for collinear momenta. Here and 
below we indicate the scaling properties of the components 
$(n\cdot p,\bar n\cdot p,p_\perp)$. The corresponding effective-theory 
fields and their scaling relations are $h_v\sim\lambda^{3/2}$ (soft heavy 
quark), $q_s\sim\lambda^{3/2}$ (soft light quark), 
$A_s^\mu\sim(\lambda,\lambda,\lambda)$ (soft gluon), and 
$\xi\sim\lambda$ (collinear quark), $A_c^\mu\sim(\lambda^2,1,\lambda)$ 
(collinear gluon). The collinear quark is described by a 2-component 
spinor subject to the constraint $\nslash\,\xi=0$.

Short-distance fluctuations in exclusive $B$ decays are usually 
characterized by two different large scales: the hard scale 
$E^2\sim m_b^2$ associated with off-shell fluctuations of the heavy 
quark, and the ``hard-collinear'' scale $p_s\cdot p_c\sim E\Lambda$ 
arising in interactions involving both soft and collinear degrees of 
freedom. In order to disentangle the physics associated with these two 
scales it is sometimes useful to perform the matching of full QCD onto 
the low-energy effective theory in two steps, by going through an 
intermediate effective theory (called SCET$_{\rm I}$) containing 
``hard-collinear'' modes with virtualities $p_{hc}^2\sim E\Lambda$ as 
dynamical degrees of freedom. In a second step SCET$_{\rm I}$ is matched 
onto the final theory (called SCET$_{\rm II}$) containing near on-shell 
soft and collinear partons only. In this paper we are concerned with the 
structure of this final effective theory.

\begin{figure}
\begin{center}
\begin{tabular}{cccc}
\psfrag{a}[b]{C}
\psfrag{b}[b]{~S}
\psfrag{c}[bl]{S}
\psfrag{d}[b]{~C}
\psfrag{e}[b]{HC}
\includegraphics[height=0.17\textwidth]{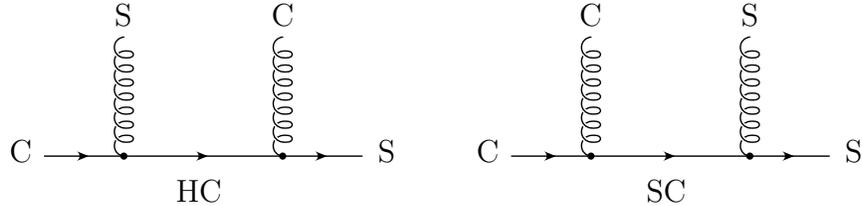} &&&
\psfrag{a}[b]{C}
\psfrag{b}[b]{~C}
\psfrag{c}[bl]{S}
\psfrag{d}[b]{~S}
\psfrag{e}[b]{SC}
\includegraphics[height=0.17\textwidth]{hardCollInt.ps}
\end{tabular}
\end{center}
\vspace{-0.8cm}
\centerline{\parbox{14cm}{\caption{\label{fig:messenger}
Examples of interactions between two soft and two collinear fields 
induced by the exchange of a hard-collinear particle (left) and of a 
soft-collinear particle (right). Hard-collinear modes are integrated out 
in SCET$_{\rm II}$, while soft-collinear modes remain as low-energy 
degrees of freedom.}}}
\end{figure}

At leading order in power counting the effective strong-interaction 
Lagrangian of SCET$_{\rm II}$ splits up into separate Lagrangians for the 
soft and collinear fields. This property implies factorization of many 
processes involving soft and collinear partons at leading power in 
$\lambda$. Factorization is however not guaranteed for quantities that 
vanish at leading power, such as heavy-to-light form factors at large 
recoil. It was argued in \cite{Hill:2002vw} that at subleading order in 
$\lambda$ interactions between soft and collinear particles occur, which 
can violate factorization. As illustrated in Figure~\ref{fig:messenger}, 
these interactions can be mediated by the exchange of a short-distance 
hard-collinear mode, or by a long-distance ``messenger particle'' with 
momentum scaling $E(\lambda^2,\lambda,\dots)$, where the transverse 
momentum components can be at most of $O(\lambda)$. We will argue in the 
present work that these two exchange mechanisms contribute under 
different kinematic conditions. In particular, the hard-collinear 
exchange requires the presence of collinear particles in the initial 
state and so is irrelevant for SCET applications to $B$ decays. The 
scaling of the long-distance messenger particle is such that it can 
couple to both soft and collinear fields without taking them far 
off-shell. It was left open in \cite{Hill:2002vw} whether the messenger 
exchange should be described in terms of a new field in the effective 
theory, or by considering the exchange particle as a soft (or collinear) 
field subject to certain constraints on some of its momentum components.

To answer this question, one must determine whether propagators with 
scaling corresponding to the messenger particles can give rise to pinch 
singularities in Feynman loop diagrams. By the Coleman--Norton theorem
such singularities only arise from on-shell intermediate states 
\cite{Coleman}, and hence the new mode would have to have momentum 
scaling $p_{sc}^\mu\sim E(\lambda^2,\lambda,\lambda^{3/2})$. Since this 
is the ``largest'' on-shell mode that can couple to both soft and 
collinear fields without affecting their momentum scaling, we call this 
mode ``soft-collinear''.  Naively, one would expect that the low-energy 
effective theory would only contain soft and collinear fields scaling in 
the same way as the external momenta of soft and collinear hadrons, 
especially since the soft-collinear momentum corresponds to a virtuality 
$p_{sc}^2\sim E^2\lambda^3$ that, for $\lambda\sim\Lambda/E$, is below 
the scale $\Lambda^2$. However, the evaluation of sample one-loop 
diagrams reveals that the situation is more complicated. The interplay of 
soft and collinear kinematics makes it necessary to introduce modes with 
virtuality $E^2\lambda$ (hard-collinear) and $E^2\lambda^3$ 
(soft-collinear) in addition to collinear and soft modes. In the 
low-energy theory the modes with off-shellness $E^2\lambda$ are 
integrated out and lead to the occurrence of operators which are smeared 
over large distances of $O(1/\Lambda)$ \cite{Hill:2002vw}, while the 
soft-collinear modes have to be kept as degrees of freedom.

The first goal of this paper is to establish, with the help of an 
explicit example, that soft-collinear modes are part of the low-energy 
effective theory for soft and collinear partons. As a result, 
SCET$_{\rm II}$ is a more complicated theory than anticipated, and the 
matching of the intermediate effective theory SCET$_{\rm I}$ onto the 
final theory SCET$_{\rm II}$ is more involved than envisioned in 
\cite{Bauer:2002aj,Chay:2003ju,Bauer:2003mg}. In the second part of the 
paper we develop SCET$_{\rm II}$ in the presence of the new modes and 
discuss some examples of the relevance of soft-collinear messenger 
exchange.

A prominent example of quantities that are sensitive to soft-collinear 
exchange graphs are heavy-to-light form factors at large recoil, for 
which soft-collinear modes can (and do) contribute at first order in 
$\lambda$. They are needed to describe the ``soft overlap'' 
contribution, which is formally the leading contribution to the form 
factors in the heavy-quark limit \cite{formfactor}. Another important 
application of the soft-collinear modes arises when one studies the 
endpoint behavior of hard-scattering kernels in QCD factorization 
theorems. The demonstration of the absence of endpoint singularities 
is an important part of factorization proofs (see, e.g., the discussion 
in \cite{Bosch:2003fc}). In the endpoint region $x\ll 1$ the scaling of 
the momentum of a collinear parton carrying longitudinal momentum 
fraction $x$ inside a fast light hadron changes from 
$E(\lambda^2,1,\lambda)$ to $E(\lambda^2,\lambda,\dots)$. Similarly, in 
the region $l_+\ll\Lambda$ the scaling of the momentum of a soft parton 
inside the $B$ meson changes from $E(\lambda,\lambda,\lambda)$ to 
$E(\lambda^2,\lambda,\dots)$. In both cases it is natural to describe 
these endpoint configurations in terms of soft-collinear fields. For 
the case of factorization for the exclusive decay $B\to K^*\gamma$ this 
will be illustrated in \cite{Kstargamma}.

\section{Relevance of the soft-collinear mode}
\label{ref:Sudakov}

\begin{figure}
\begin{center}
\psfrag{k+p}[b]{$\phantom{aaaaaa}k+p$}
\psfrag{k+l}[b]{$k+l\,$}
\psfrag{p}[b]{$\phantom{a}p$}
\psfrag{k}[b]{$~~k$}
\psfrag{l}[b]{$l$}
\includegraphics[width=0.25\textwidth]{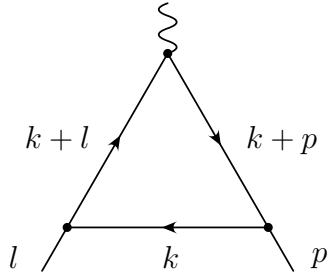}
\end{center}
\vspace{-0.4cm}
\centerline{\parbox{14cm}{\caption{\label{fig:triangle}
Scalar triangle diagram with an external soft momentum $l$ and a 
collinear momentum $p$. The loop momentum is denoted by $k$.}}}
\end{figure}

It will be instructive to demonstrate the relevance of soft-collinear 
modes with an explicit example. Consider the scalar triangle graph shown 
in Figure~\ref{fig:triangle} in the kinematic region where the external 
momenta are $l^\mu\sim(\lambda,\lambda,\lambda)$ soft, 
$p^\mu\sim(\lambda^2,1,\lambda)$ collinear, and 
$q^\mu=(l-p)^\mu\sim(\lambda,1,\lambda)$ hard-collinear. We define the 
loop integral
\begin{equation}
   I = i\pi^{-d/2} \mu^{4-d} \int d^dk\,
   \frac{1}{(k^2+i0)\,[(k+l)^2+i0]\,[(k+p)^2+i0]}
\end{equation}
in $d=4-2\epsilon$ space-time dimensions and analyze it for arbitrary 
external momenta obeying the above scaling relations. It will be 
convenient to define the invariants
\begin{equation}
   L^2\equiv -l^2 - i0 \,, \qquad
   P^2\equiv -p^2 - i0 \,, \qquad
   Q^2\equiv -(l-p)^2-i0 = 2l_+\cdot p_- - i 0 + \dots \,,
\end{equation}
which scale like $L^2,P^2\sim\lambda^2$ and $Q^2\sim\lambda$. (In 
physical units, $L^2,P^2\sim\Lambda^2$ and $Q^2\sim E\Lambda$ with 
$E\gg\Lambda$.) As long as these momenta are off-shell the integral is 
ultra-violet and infra-red finite and can be evaluated setting 
$\epsilon=0$, with the result
\begin{equation}\label{Iexact}
   I = \frac{1}{Q^2} \left[ \ln\frac{Q^2}{L^2} \ln\frac{Q^2}{P^2}
   + \frac{\pi^2}{3} + O(\lambda) \right] .
\end{equation}

Let us now try to reproduce this result by evaluating the contributions 
from different momentum modes. The method of regions 
\cite{Beneke:1997zp,Smirnov:pj} can be used to find the momentum 
configurations giving rise to leading-order contributions to the integral 
$I$. There is no hard contribution, since for $k^\mu\sim E(1,1,1)$ the 
integrand can be Taylor-expanded in the external momenta, giving 
scaleless integrals that vanish in dimensional regularization. A 
short-distance contribution arises from the region of hard-collinear loop 
momenta $k^\mu\sim E(\lambda,1,\lambda^{1/2})$, and power counting shows 
that it is indeed of leading power: 
$I_{\rm HC}\sim\lambda^2\cdot(\lambda^{-1})^3\sim\lambda^{-1}$ (where we 
display the scaling of the integration measure and of the three 
propagators), which is of the same order as the leading term in the 
result (\ref{Iexact}). Simplifying the propagators in the hard-collinear 
region we obtain at leading power
\begin{eqnarray}
   I_{\rm HC}
   &=& i\pi^{-d/2} \mu^{4-d} \int d^dk\,
    \frac{1}{(k^2+i0)\,(k^2 + 2k_-\cdot l_+ + i0)\,
             (k^2 + 2k_+\cdot p_- + i0)} \nonumber\\
   &=& \frac{\Gamma(1+\epsilon)}{2l_+\cdot p_-}\,
    \frac{\Gamma^2(-\epsilon)}{\Gamma(1-2\epsilon)}
    \left( \frac{\mu^2}{2l_+\cdot p_-} \right)^\epsilon \nonumber\\
   &=& \frac{\Gamma(1+\epsilon)}{Q^2} \left( \frac{1}{\epsilon^2}
    + \frac{1}{\epsilon} \ln\frac{\mu^2}{Q^2}
    + \frac12 \ln^2\frac{\mu^2}{Q^2} - \frac{\pi^2}{6} \right)
    + O(\epsilon) \,,
\end{eqnarray}
where in the last step we have replaced $2l_+\cdot p_-\to Q^2$, which is 
legitimate at leading power. The relevant physical scale of this 
contribution is the hard-collinear scale $\mu^2\sim Q^2\sim E\Lambda$.

Long-distance contributions to the integral arise from the regions of 
soft or collinear loop momenta, where 
$k^\mu\sim E(\lambda,\lambda,\lambda)$ or 
$k^\mu\sim E(\lambda^2,1,\lambda)$, respectively. Power counting shows 
that both regions give rise to leading-order contributions: 
$I_{\rm S}\sim \lambda^4\cdot(\lambda^{-2})^2\cdot\lambda^{-1}%
\sim\lambda^{-1}$, and $I_{\rm C}\sim\lambda^4\cdot\lambda^{-2}\cdot%
\lambda^{-1}\cdot\lambda^{-2}\sim\lambda^{-1}$. Simplifying the 
propagators in the soft region we obtain at leading power
\begin{eqnarray}
   I_{\rm S}
   &=& i\pi^{-d/2} \mu^{4-d} \int d^dk\,
    \frac{1}{(k^2+i0)\,[(k+l)^2+i0]\,(2k_+\cdot p_- + i0)} \nonumber\\
   &=& - \frac{\Gamma(1+\epsilon)}{2l_+\cdot p_-}\,
    \frac{\Gamma^2(-\epsilon)}{\Gamma(1-2\epsilon)}
    \left( \frac{\mu^2}{L^2} \right)^\epsilon \nonumber\\
   &=& \frac{\Gamma(1+\epsilon)}{Q^2} \left( - \frac{1}{\epsilon^2}
    - \frac{1}{\epsilon} \ln\frac{\mu^2}{L^2}
    - \frac12 \ln^2\frac{\mu^2}{L^2} + \frac{\pi^2}{6} \right)
    + O(\epsilon) \,.
\end{eqnarray}
The relevant physical scale of this contribution is the soft scale 
$\mu^2\sim L^2\sim\Lambda^2$. Similarly, in the collinear region we 
obtain
\begin{eqnarray}
   I_{\rm C}
   &=& i\pi^{-d/2} \mu^{4-d} \int d^dk\,
    \frac{1}{(k^2+i0)\,(2k_-\cdot l_+ + i0)\,[(k+p)^2+i0]} \nonumber\\
   &=& - \frac{\Gamma(1+\epsilon)}{2l_+\cdot p_-}\,
    \frac{\Gamma^2(-\epsilon)}{\Gamma(1-2\epsilon)}
    \left( \frac{\mu^2}{P^2} \right)^\epsilon \nonumber\\
   &=& \frac{\Gamma(1+\epsilon)}{Q^2} \left( - \frac{1}{\epsilon^2}
    - \frac{1}{\epsilon} \ln\frac{\mu^2}{P^2}
    - \frac12 \ln^2\frac{\mu^2}{P^2} + \frac{\pi^2}{6} \right)
    + O(\epsilon) \,.
\end{eqnarray}
The relevant physical scale of this contribution is the collinear scale 
$\mu^2\sim P^2\sim\Lambda^2$.

\begin{figure}
\begin{center}
\begin{tabular}{cccccccc}
\raisebox{\ph}{$\delta C\,\times$} 
\includegraphics[width=\gw]{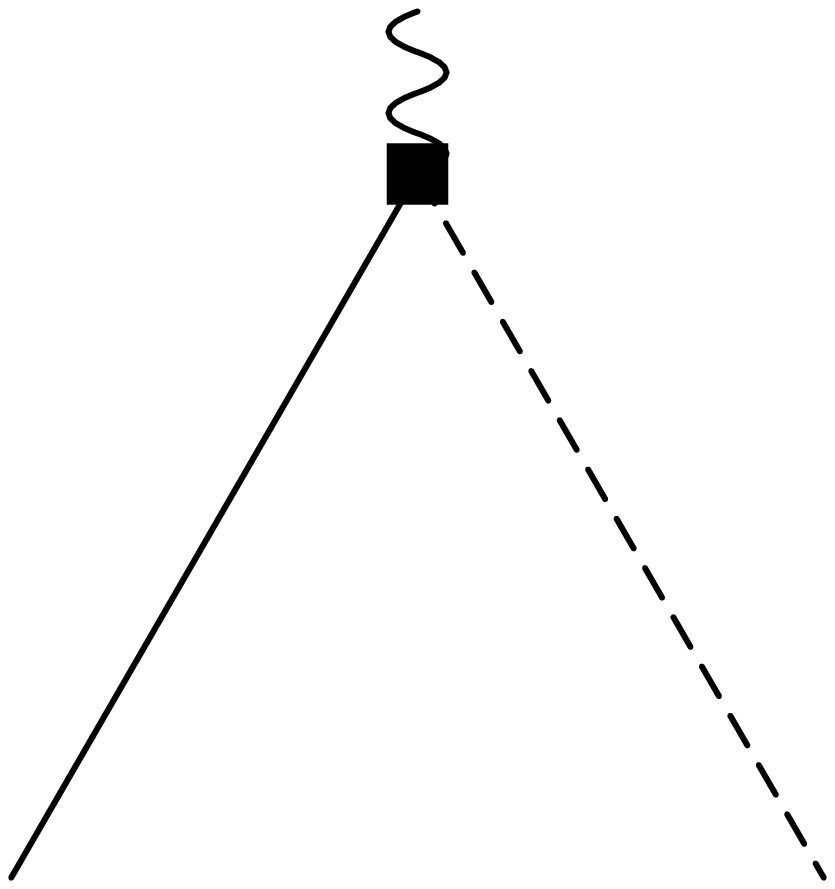}
 & \raisebox{\ph}{$+$} & \includegraphics[width=\gw]{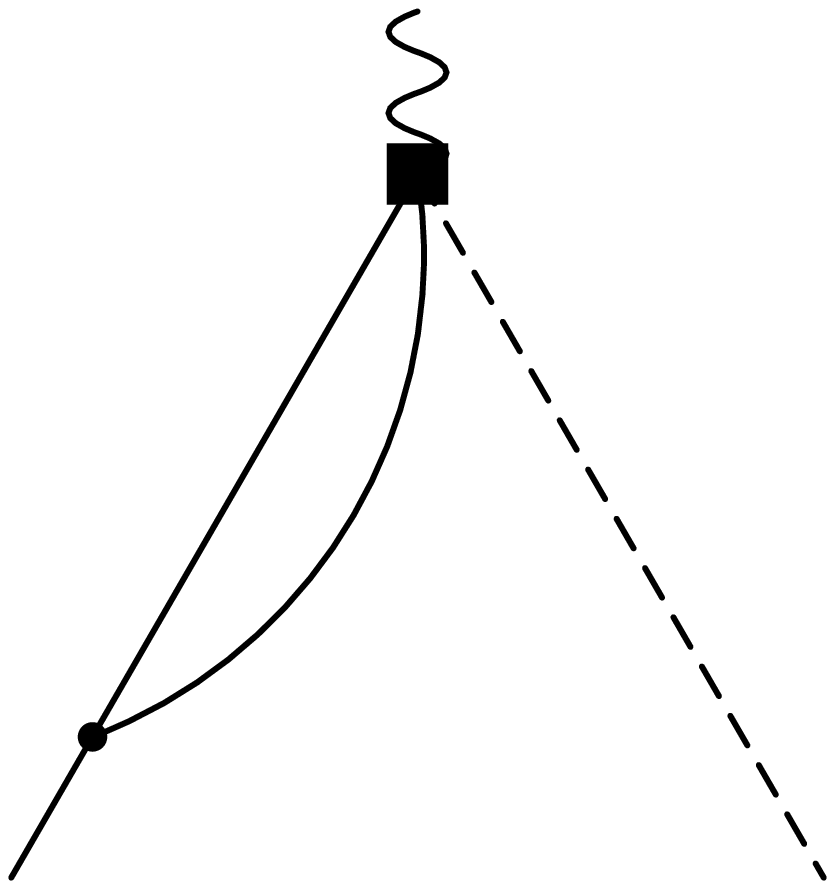}
 & \raisebox{\ph}{$+$} & \includegraphics[width=\gw]{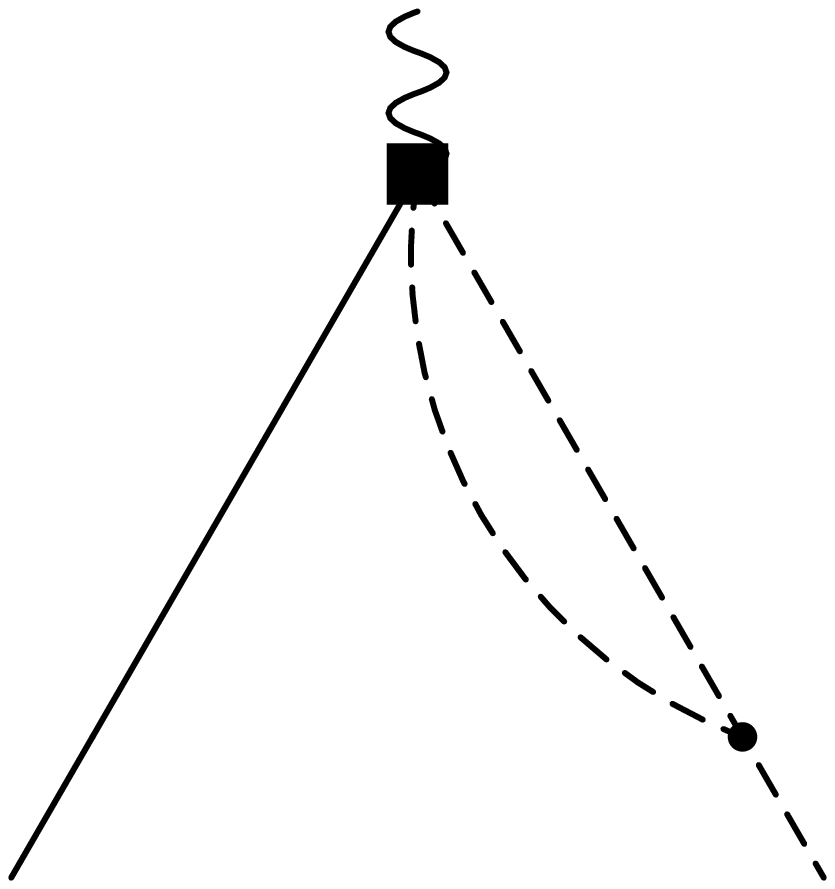}
 & \raisebox{\ph}{$+$} & \includegraphics[width=\gw]{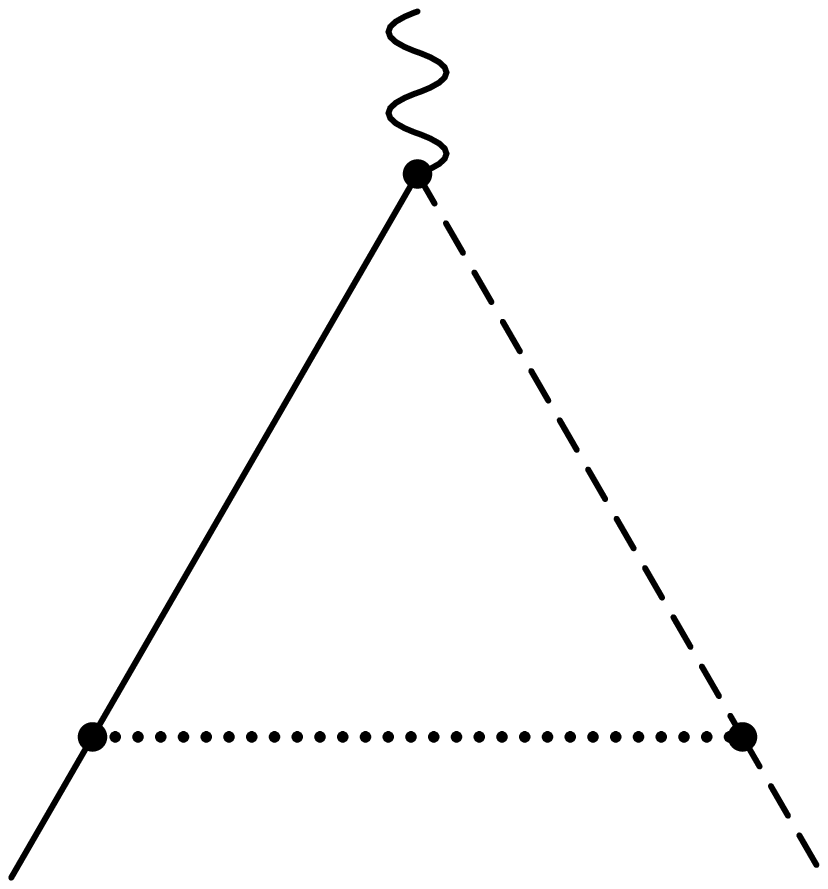}
\end{tabular}
\end{center}
\vspace{-0.2cm}
\centerline{\parbox{14cm}{\caption{\label{fig:SCETgraphs}
Effective field-theory graphs. Full lines denote soft fields, dashed 
lines collinear fields, and dotted lines soft-collinear fields.}}}
\end{figure}

Figure~\ref{fig:SCETgraphs} illustrates that the three contributions 
derived above have a representation in terms of a low-energy effective 
theory containing soft and collinear fields, in which the hard-collinear 
modes are integrated out. Here $\delta C$ denotes the hard-collinear 
contribution to the Wilson coefficient of the current 
operator containing a soft and a collinear field. This coefficient arises 
from integrating out the short-distance hard-collinear modes. In the 
second and third diagrams the hard-collinear propagators have been shrunk 
to a point, leaving loops of only soft or only collinear lines. 

The sum $I_{\rm HC}+I_{\rm C}+I_{\rm S}$ does not reproduce the exact 
leading-order term in (\ref{Iexact}). In fact, the two expressions differ 
by large single and double (Sudakov) logarithms of the form 
$\ln(\mu^2 Q^2/L^2 P^2)$, which remain large even at a low scale 
$\mu^2\sim\Lambda^2$. The discrepancy is due to the presence of another
leading region, which arises when the loop momentum scales like
$k^\mu\sim E(\lambda^2,\lambda,\lambda^{3/2})$. Power counting shows that 
this indeed gives rise to a leading-order contribution: $I_{\rm SC}\sim%
\lambda^6\cdot\lambda^{-3}\cdot(\lambda^{-2})^2\sim\lambda^{-1}$. 
Simplifying the propagators in the soft-collinear region we find
\begin{eqnarray}
   I_{\rm SC}
   &=& i\pi^{-d/2} \mu^{4-d} \int d^dk\,
    \frac{1}{(k^2+i0)\,(2k_-\cdot l_+ + l^2 + i0)\,
             (2k_+\cdot p_- + p^2 + i0)} \nonumber\\
   &=& - \frac{\Gamma(1+\epsilon)}{2l_+\cdot p_-}\,
    \Gamma(\epsilon)\,\Gamma(-\epsilon)\,
    \left( \frac{2\mu^2 l_+\cdot p_-}{L^2 P^2} \right)^\epsilon
    \nonumber\\
   &=& \frac{\Gamma(1+\epsilon)}{Q^2} \left( \frac{1}{\epsilon^2}
    + \frac{1}{\epsilon} \ln\frac{\mu^2 Q^2}{L^2 P^2}
    + \frac12 \ln^2\frac{\mu^2 Q^2}{L^2 P^2} + \frac{\pi^2}{6} \right)
    + O(\epsilon) \,.
\end{eqnarray}
The relevant scale of this contribution is a particular combination of 
the hard-collinear, soft, and collinear scales, 
$\mu^2\sim (L^2 P^2)/Q^2\sim\Lambda^3/E$, which (for 
$\lambda\sim\Lambda/E$) is parametrically smaller than the QCD scale 
$\Lambda^2$.

The soft-collinear contribution precisely accounts for the difference 
encountered above, so that 
\begin{equation}
   I = I_{\rm HC} + I_{\rm S} + I_{\rm C} + I_{\rm SC}
\end{equation}
up to higher-order terms in $\lambda$. This example demonstrates that 
soft-collinear modes are required to reproduce the correct analytic 
structure of full-theory amplitudes containing both soft and collinear 
external momenta. One must then introduce a new field in the 
low-energy effective theory, whose contribution is represented by the 
last diagram in Figure~\ref{fig:SCETgraphs}.

\begin{figure}
\begin{center}
\begin{tabular}{c}
\psfrag{l2}[b]{\small $m_b v+l_2$}
\psfrag{p1}[b]{\small $p_2$}
\psfrag{p2}[B]{\small $p_1$}
\psfrag{l1}[B]{\small $l_1$}
\psfrag{B}[B]{\phantom{,}$B$}
\psfrag{M}[B]{\phantom{a} $M$}
\includegraphics[width=0.4\textwidth]{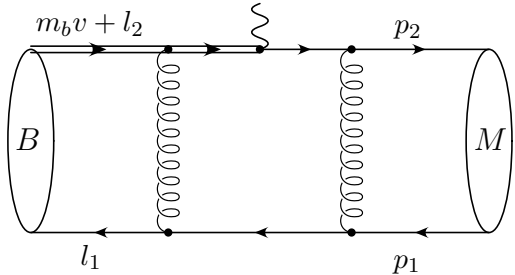}
\end{tabular}
\end{center}
\vspace{-0.2cm}
\centerline{\parbox{14cm}{\caption{\label{fig:box}
A QCD diagram contributing to the decay of a $B$ meson to an energetic 
light meson $M$. The relevant loop subgraph is a pentagon with two 
collinear external lines with momenta $p_1$, $p_2$, a soft line with 
momentum $l_1$, and a heavy-quark line with momentum $m_b v+l_2$.}}}
\end{figure}

The above conclusion is completely general and does not rely on the
particular diagram investigated here. For instance, we have analyzed the 
corresponding vertex diagram in QCD, as well as the QCD pentagon subgraph 
shown in Figure~\ref{fig:box}, which contributes to heavy-to-light form 
factors at large recoil. We again find that soft-collinear modes are 
necessary to reproduce the correct analytic structure of the diagrams, 
i.e., the QCD vertex graph receives a leading-order contribution from 
the soft-collinear exchange graph shown in the last diagram in 
Figure~\ref{fig:SCETgraphs}, and the pentagon subgraph in 
Figure~\ref{fig:box} receives a leading-order contribution from 
the region where the anti-quark line between the two lower gluon 
attachments carries a soft-collinear momentum.

\section{Relation with the Sudakov form factor}
\label{sec:Sudakov}

In order to build up intuition for the new soft-collinear mode it may be 
instructive to consider the following analogy with the off-shell Sudakov
form factor. Starting from the kinematic situation in 
Figure~\ref{fig:triangle} we can perform a longitudinal Lorentz boost 
into the Breit frame, in which the two 3-momenta $\vec{l}$ and $\vec{p}$ 
are equal in magnitude and opposite in direction. This boost rescales the
components $(n\cdot q,\bar n\cdot q,q_\perp)$ of all 4-momenta into 
$(\lambda^{-1/2}\,n\cdot q,\lambda^{1/2}\,\bar n\cdot q,q_\perp)$. 
Introducing a new expansion parameter $\hat\lambda\equiv\lambda^{1/2}$ we 
then find the following correspondence between modes in the original 
frame and in the Breit frame:
\begin{equation}
\begin{aligned}
   &\mbox{hard-collinear:} &\quad
   &E(\lambda,1,\lambda^{1/2}) 
   &\leftrightarrow\qquad
   &\mbox{hard:} &
   &\hat E(1,1,1) \\
   &\mbox{soft:} &
   &E(\lambda,\lambda,\lambda) \qquad
   &\leftrightarrow\qquad
   &\mbox{anti-collinear:} &\quad
   &\hat E(1,\hat\lambda^2,\hat\lambda) \\
   &\mbox{collinear:} &
   &E(\lambda^2,1,\lambda)
   &\leftrightarrow\qquad
   &\mbox{collinear:} &
   &\hat E(\hat\lambda^2,1,\hat\lambda) \\
   &\mbox{soft-collinear:} &
   &E(\lambda^2,\lambda,\lambda^{3/2})
   &\leftrightarrow\qquad
   &\mbox{ultra-soft:} &
   &\hat E(\hat\lambda^2,\hat\lambda^2,\hat\lambda^2) \\
\end{aligned}
\end{equation}
where $\hat E\equiv E\hat\lambda$. These are precisely the modes arising 
in the analysis of the off-shell Sudakov form factor 
\cite{Smirnov:1998vk,Kuhn:1999nn}. In the language of effective field 
theory, it follows that the original SCET$_{\rm II}$ problem (with 
expansion parameter $\lambda$ and large scale $E$) can be mapped onto a 
SCET$_{\rm I}$ problem (with expansion parameter $\hat\lambda$ and large 
scale $\hat E$) containing two types of collinear fields along with 
ultra-soft fields, which correspond to the soft-collinear messenger modes 
of the original problem.\footnote{However, this mapping cannot be done 
for processes involving heavy quarks, in which a natural Lorentz frame is 
defined by the rest frame of a heavy hadron.}  
We hope this analogy with a familiar problem will help to convince the 
reader of the relevance of soft-collinear modes in SCET$_{\rm II}$. We 
now proceed to construct the low-energy effective theory including the 
corresponding fields.

\section{The soft-collinear Lagrangian}
\label{sec:softcol}

We start by studying the scaling properties and self-interactions of 
soft-collinear fields. We introduce soft-collinear gauge and fermion 
fields, $A_{sc}^\mu$ and $q_{sc}$, in the usual way. The scaling 
properties of these fields follow from an analysis of the corresponding 
two-point functions in position space \cite{Beneke:2002ph}, taking into 
account that $p_{sc}^2\sim\lambda^3$ and $d^4p_{sc}\sim\lambda^6$. We 
find that
\begin{equation}
   A_{sc}^\mu \sim (\lambda^2,\lambda,\lambda^{3/2})
\end{equation}
scales like a soft-collinear momentum, which guarantees homogeneous 
scaling laws for the components of the soft-collinear covariant 
derivative $iD_{sc}^\mu=i\partial^\mu+A_{sc}^\mu$. Here and below, a 
factor of $g_s$ is included in the definition of the gauge fields. 

The fermion field can be split up into large and small components with 
different scaling relations. We define $q_{sc}=\theta+\sigma$, where 
\begin{equation}
   \theta = \frac{\nslash\nbslash}{4}\,q_{sc} \,, \qquad
   \sigma = \frac{\nbslash\nslash}{4}\,q_{sc} \,, \qquad
   \mbox{with} \quad \nslash\theta = \nbslash\sigma = 0 \,.
\end{equation}
The analysis of the fermion two-point function reveals that 
\begin{equation}
   \theta\sim\lambda^2 \,, \qquad \sigma\sim\lambda^{5/2} \,.
\end{equation}

As long as we consider interactions of only soft-collinear fields, 
nothing prevents us from boosting to a Lorentz frame in which these 
fields have homogeneous momentum scaling 
$p^\mu\sim E(\lambda^{3/2},\lambda^{3/2},\lambda^{3/2})$. This is 
analogous to the case of the collinear Lagrangian, which in this sense is 
equivalent to the ordinary QCD Lagrangian. It follows that the effective 
Lagrangian for soft-collinear fields has the same form as the collinear 
Lagrangian \cite{Bauer:2000yr,Beneke:2002ph}, i.e.\
\begin{equation}\label{Lsc}
   {\cal L}_{sc} = \bar\theta\,\frac{\nbslash}{2}\,in\cdot D_{sc}\,\theta
   - \bar\theta\,i\Dslash_{sc\perp}\,\frac{\nbslash}{2}\,
   \frac{1}{i\bar n\cdot D_{sc}}\,i\Dslash_{sc\perp}\,\theta \,.
\end{equation}
To obtain this result one simply inserts the decomposition 
$q_{sc}=\theta+\sigma$ into the Dirac Lagrangian and eliminates the 
small-component field $\sigma$ using its equation of motion, which yields 
\begin{equation}\label{sigmasol}
   \sigma = - \frac{\nbslash}{2}\,\frac{1}{i\bar n\cdot D_{sc}}\,
   i\Dslash_{sc\perp}\,\theta \,.
\end{equation}
It is straightforward to check that the operators in the effective
Lagrangian (\ref{Lsc}) scale like $\lambda^6$, which when combined with 
the scaling of the soft-collinear measure $d^4x\sim\lambda^{-6}$ ensures 
that these terms are of leading order in power counting. The fermion 
Lagrangian given above must be complemented by the pure-gauge and ghost 
Lagrangians, which retain the same form as in ordinary QCD. Using 
arguments along the lines of \cite{Beneke:2002ph} it can be shown that 
the Lagrangian ${\cal L}_{sc}$ is not renormalized. Below, we will often 
write expressions in terms of the two components $\theta$ and $\sigma$, 
keeping in mind that at the end $\sigma$ may be eliminated using 
(\ref{sigmasol}).

\section{Interactions of soft-collinear fields with other fields}

The effective Lagrangian of SCET$_{\rm II}$ can be derived by decomposing 
the QCD fields into the various modes and integrating out the hard 
and hard-collinear modes. Here we focus on the pure QCD Lagrangian 
without external operators mediating weak interactions. In general, the 
effective Lagrangian can be split up as
\begin{equation}\label{Lscet}
   {\cal L}_{\rm SCET_{\rm II}}
   = {\cal L}_s + {\cal L}_c + {\cal L}_{sc}
   + {\cal L}_{s+sc}^{\rm int} + {\cal L}_{c+sc}^{\rm int}
   ~\,[\, + {\cal L}_{s+c}^{\rm int} \,] \,,
\end{equation}
where the first three terms correspond to the Lagrangians of soft 
particles (including heavy quarks), collinear particles, and 
soft-collinear particles. The term ${\cal L}_{s+c}^{\rm int}$ in 
brackets corresponds to effective interactions among soft and collinear 
particles induced by the exchange of hard-collinear modes, which arise 
at subleading order in $\lambda$ \cite{Hill:2002vw}. We will argue in 
Section~\ref{sec:induced} that these interactions are kinematically 
forbidden in $B$ decays. This term can thus be dropped from the SCET 
Lagrangian. Our focus in this paper is on the terms 
${\cal L}_{s+sc}^{\rm int}$ and ${\cal L}_{c+sc}^{\rm int}$, describing 
the interactions involving soft-collinear fields. Note that the 
integration measures $d^4x$ in the action 
$S_{\rm SCET_{\rm II}}=\int d^4x\,{\cal L}_{\rm SCET_{\rm II}}$ scale 
differently for the various terms above. The measures for ${\cal L}_s$ 
and ${\cal L}_c$ scale like $\lambda^{-4}$, while that for 
${\cal L}_{sc}$ scales like $\lambda^{-6}$. The measure for the 
interaction Lagrangian ${\cal L}_{s+c}^{\rm int}$ scales like 
$\lambda^{-3}$, whereas those for the interaction Lagrangians 
${\cal L}_{s+sc}^{\rm int}$ and ${\cal L}_{c+sc}^{\rm int}$ scale like 
$\lambda^{-4}$ (see below).

Soft-collinear fields can couple to collinear or soft fields without 
altering their scaling properties. Momentum conservation implies that in 
QCD (i.e., at the level of three- and four-point vertices) soft-collinear 
fields can only couple to either soft or collinear modes, but not both. 
In such interactions more than one soft or collinear particle must be 
involved. These ``pure QCD'' interactions are always near on-shell and do 
not involve the exchange of hard or hard-collinear modes, so that we can 
perform the construction of the effective theory at tree level. The three 
relevant regions are collinear, soft, and soft-collinear. We thus split 
up the gluon field as $A^\mu=A^\mu_c+A^\mu_s+A^\mu_{sc}$ and choose a 
similar decomposition for the quark field. The QCD Lagrangian is then 
expanded in these fields, dropping terms that are forbidden by momentum 
conservation. Due to the above form of the Lagrangian we can separately 
construct the effective theory in the $(s+sc)$ and $(c+sc)$ sectors. The 
resulting effective interaction Lagrangians ${\cal L}_{s+sc}^{\rm int}$ 
and ${\cal L}_{c+sc}^{\rm int} $ are not renormalized. 

An important remark is that the soft-collinear fields in interactions 
with soft or collinear fields must be multipole expanded in order to 
properly separate the contributions from the different momentum regions 
and to avoid double counting \cite{Beneke:2002ph}. Consider a term in 
the action containing some collinear fields and some soft-collinear 
fields, e.g.\
\begin{equation}
   \int d^4x\,\phi_c(x)\,\phi_c(x)\,\phi_{sc}(x) 
   \sim \int d^4x\,e^{i(p+p'+q)\cdot x} \,,
\end{equation}
where $p,p'$ are collinear momenta and $q$ is a soft-collinear momentum. 
Integration over $d^4x$ in the action gives rise to $\delta$-functions, 
which must be used to eliminate one of the collinear momenta, not the 
soft-collinear momentum. These $\delta$-functions scale like 
$\lambda^{-4}$, which is thus the scaling of the measure $d^4x$. The 
vector $x^\mu$ scales as appropriate for the argument of a 
collinear field, i.e., 
$x^\mu\sim(1,\lambda^{-2},\lambda^{-1})$. The soft-collinear scaling
$q^\mu\sim(\lambda^2,\lambda,\lambda^{3/2})$ then implies that we can 
expand $e^{iq\cdot x}=e^{iq_+\cdot x_-}(1+iq_\perp\cdot x_\perp+\dots)$. 
For the soft-collinear field this implies the multipole expansion
\begin{equation}
   \phi_{sc}(x) = \phi_{sc}(x_-) + x_\perp\cdot\partial_\perp\,
   \phi_{sc}(x_-) + \dots \qquad \mbox{(in collinear interactions)} \,,
\end{equation}
where $x_-^\mu=\frac12(\bar n\cdot x)\,n^\mu$. The first correction 
term is of $O(\lambda^{1/2})$, and the omitted terms are of $O(\lambda)$ 
and higher. Similarly, if the soft-collinear field interacts with soft 
fields the vector $x^\mu$ scales as appropriate for a soft momentum, 
i.e., $x^\mu\sim(\lambda^{-1},\lambda^{-1},\lambda^{-1})$, and so only 
the dependence of the soft-collinear field on $x_+$ must be kept, i.e.
\begin{equation}
   \phi_{sc}(x) = \phi_{sc}(x_+) + x_\perp\cdot\partial_\perp\,
   \phi_{sc}(x_+) + \dots \qquad \mbox{(in soft interactions)} \,,
\end{equation}
where $x_+^\mu=\frac12(n\cdot x)\,\bar n^\mu$.

Soft-collinear fields can couple to soft or collinear fields without 
altering their scaling properties. This motivates the treatment of the 
soft-collinear gluon field as a background field, which is smoother 
than soft and collinear fields \cite{Beneke:2002ph}. It is also 
convenient to choose fermion field variables which do not mix under 
gauge transformations. Thus under soft gauge transformations $U_s$ the 
soft fields transform as
\begin{equation}\label{gauge1}
   A_s^\mu \to U_s\,A_s^\mu\,U_s^\dagger + U_s\,[iD_{sc}^\mu,U_s^\dagger]
    \,, \qquad
   q_s \to U_s\,q_s \,, 
\end{equation}
while the collinear and soft-collinear fields remain invariant. 
Likewise, under collinear gauge transformations $U_c$ the collinear 
fields transform as
\begin{equation}\label{gauge2}
   A_c^\mu \to U_c\,A_c^\mu\,U_c^\dagger + U_c\,[iD_{sc}^\mu,U_c^\dagger]
    \,, \qquad
   \xi \to U_c\,\xi \,,
\end{equation}
while the soft and soft-collinear fields remain invariant. Finally, under 
soft-collinear gauge transformations $U_{sc}$ the fields transform as
\begin{equation}\label{Usc}
\begin{aligned}
   A_c^\mu &\to U_{sc}\,A_c^\mu\,U_{sc}^\dagger \,,
    &\xi &\to U_{sc}\,\xi \,, \\
   A_s^\mu &\to U_{sc}\,A_s^\mu\,U_{sc}^\dagger \,,
    &q_s &\to U_{sc}\,q_s \,, \\
   A_{sc}^\mu &\to U_{sc}\,A_{sc}^\mu\,U_{sc}^\dagger
    + U_{sc}\,[i\partial^\mu,U_{sc}^\dagger] \,, \qquad
    &q_{sc} &\to U_{sc}\,q_{sc} \,.
\end{aligned}
\end{equation}
It can be seen from these relations that the combination 
$(A_s^\mu+A_{sc}^\mu)$ transforms in the usual way under both soft and 
soft-collinear gauge transformations, while $(A_c^\mu+A_{sc}^\mu)$ 
transforms in the usual way under both collinear and soft-collinear gauge 
transformations. The sum of the corresponding fermion fields, however, do
not transform as the QCD fermion field. Our strategy will therefore be to 
adopt a particular gauge in the soft and collinear sectors of the theory, 
restoring gauge invariance at a later stage. Specifically, we adopt soft 
light-cone gauge $n\cdot A_s=0$ (SLCG) and collinear light-cone gauge
$\bar n\cdot A_c=0$ (CLCG).

Another subtlety related to the implementation of the gauge 
transformations (\ref{gauge1})--(\ref{Usc}) is that they change the 
scaling behavior of the various fields, because the components of the 
covariant derivative $D_{sc}^\mu$ acting on soft or collinear fields do 
not have homogeneous scaling behavior, and also because the 
soft-collinear fields multiplying soft or collinear fields are not 
multipole expanded. In the following we discuss how to set up homogeneous 
gauge transformations that preserve the power counting of the fields, 
following closely the treatment in \cite{Beneke:2002ni}. This discussion 
is necessarily rather technical. The reader not interested in the details 
of the derivation may directly consult the final results given in 
(\ref{final}) and (\ref{finalc}).

\subsection{Interactions between soft-collinear fields and soft fields}
\label{sec:soft}

We start with the sector of the theory involving soft and soft-collinear 
fields. In order to have a well-defined power counting, we replace the 
transformation rules for soft fields in (\ref{gauge1}) and (\ref{Usc}) by 
the homogeneous gauge transformations
\begin{equation}
\begin{aligned}
   \mbox{soft:}
    \quad &
    n\cdot A_s\to U_s\,n\cdot A_s\,U_s^\dagger 
    + U_s\,[in\cdot\partial,U_s^\dagger] \,, 
    & & q_s\to U_s\,q_s \,, & \\
    \quad &
    A_{s\perp}^\mu\to U_s\,A_{s\perp}^\mu\,U_s^\dagger 
    + U_s\,[i\partial_\perp^\mu,U_s^\dagger] \,, & \\
    \quad & 
    \bar n\cdot A_s\to U_s\,\bar n\cdot A_s\,U_s^\dagger 
    + U_s\,[i\bar n\cdot D_{sc}(x_+),U_s^\dagger] \,, & \\
   \mbox{soft-collinear:}
    \quad & A_s^\mu\to U_{sc}(x_+)\,A_s^\mu\,
    U_{sc}^\dagger(x_+) \,,
    & & q_s\to U_{sc}(x_+)\,q_s \,,
\end{aligned} 
\end{equation}
which are obtained by consistently keeping the leading-order terms in
each of the original transformation rules. The soft-collinear fields 
transform in the same way as before. Here and below we use the notation
that fields without argument live at position $x$, whereas some of the
soft-collinear fields live at position $x_+$ as indicated.

The new soft quark and gluon fields obeying the homogeneous 
transformation rules are related to the original ones by a 
(field-dependent, non-linear, and non-local) field redefinition. As shown 
in \cite{Beneke:2002ni}, soft fields $\hat q_s$ and $\hat A_s^\mu$ having 
these gauge transformations are given by
\begin{equation}\label{newfields}
\begin{aligned}
   q_s \big|_{\rm SLCG}
   &= R_s\,S_s^\dagger\,\hat q_s \,, \\
   A_{s\perp}^\mu \big|_{\rm SLCG}
   &= R_s\,S_s^\dagger\,(i\hat D_{s\perp}^\mu\,S_s)\,R_s^\dagger \,, \\
   \bar n\cdot A_s \big|_{\rm SLCG}
   &= R_s\,\Big[ S_s^\dagger\,(i\bar n\cdot\hat D_s\,S_s)
    + S_s^\dagger\,\bar n\cdot A_{sc}(x_+)\,S_s
    - \bar n\cdot A_{sc}(x_+) \Big]\,R_s^\dagger \,, 
\end{aligned}
\end{equation}
where the fields on the left-hand side are in soft light-cone gauge. The 
quantity
\begin{equation}
   S_s(x) \equiv \mbox{P}\exp\left( i\int_{-\infty}^0 dt\,
   n\cdot\hat A_s(x+tn) \right) 
\end{equation}
with $t\sim\lambda^{-1}$ is a soft Wilson line (expressed in terms of the 
new gluon field) along the $n$-direction, and
\begin{equation}
   R_s(x) = \mbox{P}\exp\left( i\int_0^1\!dt\,(x-x_+)_\mu\,
   A_{sc}^\mu(x_+ + t(x-x_+)) \right)
\end{equation}
is the gauge string of soft-collinear fields from $x_+$ to $x$. This 
quantity differs from 1 by terms of $O(\lambda^{1/2})$ and so can be 
expanded; this will be used below. Note that $S_s$ has the simple 
transformation properties
\begin{equation}\label{Strafo}
   S_s \to U_s\,S_s \,, \qquad
   S_s \to U_{sc}(x_+)\,S_s\,U_{sc}^\dagger(x_+) \,,
\end{equation} 
because the arguments of the soft fields in the path-ordered exponential
correspond to the same $x_+$. The quantity $R_s$ is invariant under
soft gauge transformations and transforms like
\begin{equation}\label{Rtrafo}
   R_s(x) \to U_{sc}(x)\,R_s(x)\,U_{sc}^\dagger(x_+)
\end{equation}
under soft-collinear gauge transformations. It follows that the 
expressions on the right-hand side of (\ref{newfields}) are invariant 
under soft gauge transformations and transform as ordinary QCD quark and
gluon fields (at position $x$, not $x_+$) under soft-collinear 
gauge transformations. The interpretation of (\ref{newfields}) is that 
the gauge transformation $S_s$ puts the hatted fields in soft light-cone 
gauge and the $R_s$ transformation ``de-homogenizes'' them, i.e., it 
converts the fields with homogeneous transformation laws into fields 
satisfying ordinary gauge transformations.

To obtain the effective Lagrangian ${\cal L}_{s+sc}^{\rm int}$ in 
(\ref{Lscet}) we adopt soft light-cone gauge and insert the decomposition 
$q=q_s+q_{sc}$ into the Dirac Lagrangian, dropping terms that are 
forbidden by momentum conservation. This yields
\begin{equation}
   \bar q\,i\Dslash\,q \big|_{\rm SLCG}
   \to \bar q_s\,i\Dslash_{s+sc}\,q_s
   + \bar q_s\,\Aslash_s\,q_{sc} + \bar q_{sc}\,\Aslash_s\,q_s
   + \bar q_{sc}\,i\Dslash_{sc}\,q_{sc} \,.
\end{equation}
After elimination of the small-component field $\sigma$ the last term 
gives rise to the soft-collinear Lagrangian discussed in 
Section~\ref{sec:softcol}. Let us then focus on the remaining terms 
and express them in terms of the homogenized fields defined in 
(\ref{newfields}). After a straightforward calculation we find that
\begin{eqnarray}\label{step1}
   {\cal L}_{\rm quark}
   &\to& \bar{\hat q}_s \left( i\hat\Dslash_s
    + \frac{\nslash}{2}\,\bar n\cdot A_{sc}(x_+) \right) \hat q_s
    + \bar{\hat q}_s\,S_s \left(
    R_s^\dagger\,i\Dslash_{sc}\,R_s - i\delslash
    - \frac{\nslash}{2}\,\bar n\cdot A_{sc}(x_+) \right) 
    S_s^\dagger\,\hat q_s \nonumber\\
   &+& \bigg\{ \bar{\hat q}_s\,S_s\,\frac{\nslash}{2} 
    \Big[ S_s^\dagger\,(i\bar n\cdot\hat D_s\,S_s)
    + S_s^\dagger\,\bar n\cdot A_{sc}(x_+)\,S_s
    - \bar n\cdot A_{sc}(x_+) \Big]\,R_s^\dagger\,q_{sc} \nonumber\\
   &&\mbox{}+ \bar{\hat q}_s\,(i\hat\Dslash_{s\perp}\,S_s)\,
    R_s^\dagger\,q_{sc}   + \mbox{h.c.} \bigg\} \,.
\end{eqnarray}
From now on we will drop the ``hat'' on the redefined soft fields.

To put this Lagrangian in a useful form we expand the various quantities 
involving $R_s$ in powers of $\lambda^{1/2}$ \cite{Beneke:2002ni}. We 
need
\begin{equation}\label{Rexp}
\begin{aligned}
   R_s^\dagger(x)\,q_{sc}(x)
   &= \theta(x_+) + \sigma(x_+)
    + x_{\perp}\cdot D_{sc}(x_+)\,\theta(x_+) + O(\lambda^3) \,, \\
   R_s^\dagger\,i\Dslash_{sc}\,R_s 
   &- i\delslash - \frac{\nslash}{2}\,\bar n\cdot A_{sc}(x_+)
    = \frac{\nslash}{2}\,x_{\perp\mu}\bar n_\nu\,g_s G_{sc}^{\mu\nu}(x_+)
    + O(\lambda^2) \,.
\end{aligned}
\end{equation}
Substituting these expansions into (\ref{step1}) we obtain 
\begin{eqnarray}\label{step2}
   {\cal L}_{\rm quark}
   &\to& \bar q_s \left( i\Dslash_s
    + \frac{\nslash}{2}\,\bar n\cdot A_{sc} \right) q_s
    + \bar q_s\,S_s\,\frac{\nslash}{2}\,x_{\perp\mu}\bar n_\nu\,
    g_s G_{sc}^{\mu\nu}\,S_s^\dagger\,q_s + O(\lambda^5) \nonumber\\
   &+& \bigg\{ \bar q_s\,S_s\,\frac{\nslash}{2}\,\Big( S_s^\dagger\,
    [i\bar n\cdot D_s + \bar n\cdot A_{sc}]\,S_s - i\bar n\cdot D_{sc}
    \Big)\,\sigma \nonumber\\
   &&\mbox{}+ \bar q_s\,(i\Dslash_{s\perp}\,S_s)\, 
    \Big[ (1 + x_\perp\cdot D_{sc})\,\theta + \sigma \Big] 
    + \mbox{h.c.} + O(\lambda^{11/2}) \bigg\} \,,
\end{eqnarray}
where it is now understood that all soft-collinear fields are evaluated
at position $x_+$ (after derivatives have been taken). 

The first term in the first line in the above result contains a 
leading-order interaction of soft quarks with the soft-collinear gluon 
field $\bar n\cdot A_{sc}$. This term can be removed by making another 
redefinition of the soft fields, which is analogous to the decoupling of 
ultra-soft gluon fields at leading power in SCET$_{\rm I}$ 
\cite{Bauer:2001yt}. We define
\begin{equation}\label{another}
  q_s(x) = W_{sc}(x_+)\,q_s^{(0)}(x) \,, \qquad
  A_s^\mu(x) = W_{sc}(x_+)\,A_s^{(0)\mu}(x)\,
  W_{sc}^\dagger(x_+) \,,
\end{equation}
where 
\begin{equation}\label{Zdef}
   W_{sc}(x_+) = \mbox{P}\exp\left( i\int_{-\infty}^0\!dt\,
   \bar n\cdot A_{sc}(x_+ + t\bar n) \right)
\end{equation}
with $t\sim\lambda^{-1}$. This object is invariant under soft and 
collinear gauge transformations, while under a soft-collinear gauge 
transformation
\begin{equation}
   W_{sc}(x_+)\to U_{sc}(x_+)\,W_{sc}(x_+) \,.
\end{equation}
Consequently, the new fields with ``(0)'' superscripts are invariant 
under soft-collinear gauge transformations, and there is no longer a 
soft-collinear background field in their transformation rules. In terms 
of these fields the first term in (\ref{step2}) reduces to the soft 
Lagrangian ${\cal L}_s=\bar q_s^{(0)} i\Dslash_s^{(0)} q_s^{(0)}$ in 
(\ref{Lscet}). The remaining terms yield contributions to the 
interaction Lagrangian ${\cal L}_{s+sc}^{\rm int}$ in (\ref{Lscet}). 
Using that $S_s=W_{sc}(x_+)\,S_s^{(0)}\,W_{sc}^\dagger(x_+)$ under the 
transformation (\ref{another}) we obtain
\begin{eqnarray}
   {\cal L}_{s+sc}^{\rm int}
   &=& \bar q_s^{(0)}\,S_s^{(0)}\,\frac{\nslash}{2}\,W_{sc}^\dagger\,
    x_{\perp\mu}\bar n_\nu\,g_s G_{sc}^{\mu\nu}\,W_{sc}\,
    S_s^{(0)\dagger}\,q_s^{(0)} + O(\lambda^5) \nonumber\\
   &+& \bigg\{ \bar q_s^{(0)}\,(i\Dslash_{s\perp}^{(0)}\,S_s^{(0)})\,
    W_{sc}^\dagger\,\Big[ (1+x_\perp\cdot D_{sc})\,\theta + \sigma \Big]
    + \bar q_s^{(0)}\,\frac{\nslash}{2}\,(i\bar n\cdot D_s^{(0)}\,
    S_s^{(0)})\,W_{sc}^\dagger\,\sigma \nonumber\\
   &&\mbox{}+ \mbox{h.c.} + O(\lambda^{11/2}) \bigg\} \,.
\end{eqnarray}
This result can be simplified further by introducing the gauge-invariant 
building blocks \cite{Hill:2002vw}
\begin{equation}
\begin{aligned}
   \Q_s &= S_s^{(0)\dagger}\,q_s^{(0)}
    = W_{sc}^\dagger(x_+)\,S_s^\dagger\,q_s \,, \\
   \A_s^\mu &= S_s^{(0)\dagger}\,(iD_s^{(0)\mu}\,S_s^{(0)}) \\[-0.15cm]
   &= W_{sc}^\dagger(x_+)\,\Big[ S_s^\dagger\,(iD_s^\mu\,S_s)
    + \frac{n^\mu}{2} \left( S_s^\dagger\,\bar n\cdot A_{sc}(x_+)\,S_s
    - \bar n\cdot A_{sc}(x_+) \right) \Big]\,W_{sc}(x_+) \,,
\end{aligned}
\end{equation}
which are invariant under both soft and soft-collinear gauge 
transformations. In terms of these fields we find the final result
\begin{eqnarray}\label{final}
   {\cal L}_{s+sc}^{\rm int}
   &=& \bar\Q_s\,\frac{\nslash}{2}\,W_{sc}^\dagger\,
    x_{\perp\mu}\bar n_\nu\,g_s G_{sc}^{\mu\nu}\,W_{sc}\,\Q_s
    + O(\lambda^5) \\
   &+& \bigg\{ \bar\Q_s\,\calAslash_{s\perp}\,W_{sc}^\dagger\,
    \Big[ (1 + x_\perp\cdot D_{sc})\,\theta + \sigma \Big] 
    + \bar\Q_s\,\frac{\nslash}{2}\,\bar n\cdot\A_s\,
    W_{sc}^\dagger\,\sigma + \mbox{h.c.} + O(\lambda^{11/2}) \bigg\} \,.
    \nonumber
\end{eqnarray}
Recall that all components of $x^\mu$ in this Lagrangian scale like 
$\lambda^{-1}$. Soft fields live at position $x$, while soft-collinear 
fields must be evaluated at position $x_+$. The small-component field 
$\sigma$ may be eliminated using (\ref{sigmasol}). Note that the 
soft-collinear fields enter this result in combinations such as 
$W_{sc}^\dagger\,\theta$, which are explicitly gauge invariant.

The measure $d^4x$ relevant to these interaction terms scales like 
$\lambda^{-4}$. It follows that in terms of the redefined fields the 
interaction of two soft quarks with a soft-collinear gluon (first line) 
is a subleading effect, for which we have computed the $
O(\lambda^{1/2})$ contribution to the action. The interaction of a soft 
quark and soft gluon with a soft-collinear quark is also a subleading 
effect, for which we have computed the $O(\lambda^{1/2})$ and 
$O(\lambda)$ contributions to the action.

In addition to the quark terms shown above there exist pure-glue 
interactions between soft-collinear and soft gluons. It can be readily 
seen that after the decoupling transformation they are also of 
subleading order in power counting. The reason is that only the 
component $\bar n\cdot A_{sc}$ of the soft-collinear field is as large 
as the corresponding component $\bar n\cdot A_s$ of the soft field. 
Since the measure $d^4x$ associated with ${\cal L}_{s+sc}^{\rm int}$ 
is the same as that for purely soft interactions, leading-order 
couplings of soft and soft-collinear gluons can only contain the 
component $\bar n\cdot A_{sc}$ of the soft-collinear gluon field. 
However, all such interactions disappear when the soft fields are 
redefined as in (\ref{another}), because
\begin{equation}
   iD_{s+sc}^\mu = iD_s^\mu + \frac{n^\mu}{2}\,\bar n\cdot A_{sc}(x_+)
   + \ldots
   = W_{sc}(x_+)\,iD_s^{(0)\mu}\,W_{sc}^\dagger(x_+) + \dots \,,
\end{equation}
where the dots denote higher-order terms in $\lambda$. The remaining 
subleading interactions between soft and soft-collinear gluons are of 
lesser phenomenological importance than the terms in (\ref{final}). 
Their precise form will not be derived here.

\subsection{Interactions between soft-collinear fields and collinear 
fields}

The discussion of the sector of the theory involving collinear and
soft-collinear fields proceeds in an analogous way. In this case we 
replace the transformation rules for collinear fields in (\ref{gauge2}) 
and (\ref{Usc}) by the homogeneous gauge transformations 
\begin{equation}
\begin{aligned}
   \mbox{collinear:}
    \quad &
    \bar n\cdot A_c\to U_c\,\bar n\cdot A_c\,U_c^\dagger 
    + U_c\,[i\bar n\cdot\partial,U_c^\dagger] \,, 
    & & \xi\to U_c\,\xi \,, & \\
    \quad &
    A_{c\perp}^\mu\to U_c\,A_{c\perp}^\mu\,U_c^\dagger 
    + U_c\,[i\partial_\perp^\mu,U_c^\dagger] \,, & \\
    \quad & 
    n\cdot A_c\to U_c\,n\cdot A_c\,U_c^\dagger 
    + U_c\,[i n\cdot D_{sc}(x_-),U_c^\dagger] \,, & \\
   \mbox{soft-collinear:}
    \quad & A_c^\mu\to U_{sc}(x_-)\,A_c^\mu\,
    U_{sc}^\dagger(x_-) \,,
    & & \xi\to U_{sc}(x_-)\,\xi \,.
\end{aligned} 
\end{equation}
Once again the soft-collinear fields transform in the same way as 
before. The new collinear quark and gluon fields obeying the homogeneous 
transformation rules are related to the original fields in collinear 
light-cone gauge $\bar n\cdot A_c=0$ by the field redefinitions 
\cite{Beneke:2002ni}
\begin{equation}
\begin{aligned}
   \xi \big|_{\rm CLCG}
   &= R_c\,W_c^\dagger\,\hat\xi \,, \\
   A_{c\perp}^\mu \big|_{\rm CLCG}
   &= R_c\,W_c^\dagger\,(i\hat D_{c\perp}^\mu\,W_c)\,R_c^\dagger \,, \\
   n\cdot A_c \big|_{\rm CLCG}
   &= R_c\,\Big[ W_c^\dagger\,(in\cdot\hat D_c\,W_c)
    + W_c^\dagger\,n\cdot A_{sc}(x_-)\,W_c
    - n\cdot A_{sc}(x_-) \Big]\,R_c^\dagger \,, 
\end{aligned}
\end{equation}
where
\begin{equation}
   W_c(x) \equiv \mbox{P}\exp\left( i\int_{-\infty}^0 dt\,
   \bar n\cdot\hat A_c(x+t\bar n) \right) 
\end{equation}
with $t\sim 1$ is a collinear Wilson line (expressed in terms of the new 
gluon field) along the $\bar n$-direction, and
\begin{equation}
   R_c(x) = \mbox{P}\exp\left( i\int_0^1\!dt\,(x-x_-)_\mu\,
   A_{sc}^\mu(x_- + t(x-x_-)) \right)
\end{equation}
is the gauge string of soft-collinear fields from $x_-$ to $x$. The
transformation properties of these objects are analogous to those for the 
Wilson lines $S_s$ and $R_s$ introduced in (\ref{Strafo}) and 
(\ref{Rtrafo}). 

To obtain the effective Lagrangian we adopt collinear light-cone gauge
and insert the decomposition $q=\xi+\eta+q_{sc}$ into the Dirac 
Lagrangian, dropping terms that are forbidden by momentum conservation. 
The only difference with respect to the discussion in 
Section~\ref{sec:soft} is that the small-component field $\eta$, which is 
part of the QCD collinear fermion field, must later be eliminated using 
its equation of motion. We then expand the terms involving the object 
$R_c$ in analogy with (\ref{Rexp}), and finally redefine the collinear 
fields in analogy with (\ref{another}), i.e.\
\begin{equation}
   \xi(x) = S_{sc}(x_-)\,\xi^{(0)}(x) \,, \quad
   \eta(x) = S_{sc}(x_-)\,\eta^{(0)}(x) \,, \quad
   A_c^\mu(x) = S_{sc}(x_-)\,A_c^{(0)\mu}(x)\,S_{sc}^\dagger(x_-) \,,
\end{equation}
where
\begin{equation}
   S_{sc}(x_-) = \mbox{P}\exp\left( i\int_{-\infty}^0\!dt\,
   n\cdot A_{sc}(x_- + tn) \right)
\end{equation}
with $t\sim\lambda^{-2}$ is defined in a similar way as the object 
$W_{sc}$ in (\ref{Zdef}). In terms of the new fields the Dirac Lagrangian
splits up into the collinear Lagrangian ${\cal L}_c$ in (\ref{Lscet}) and
terms that contribute to the interaction Lagrangian 
${\cal L}_{c+sc}^{\rm int}$ in (\ref{Lscet}). (Since we have separately
analyzed the $(s+sc)$ and $(c+sc)$ sectors we also obtain another copy 
of the Lagrangian ${\cal L}_{sc}$, which must be dropped.) These terms 
are given by
\begin{eqnarray}
   {\cal L}_{c+sc}^{\rm int}
   &=& \bar\xi^{(0)}\,W_c^{(0)}\,\frac{\nbslash}{2}\,S_{sc}^\dagger\,
    x_{\perp\mu} n_\nu\,g_s G_{sc}^{\mu\nu}\,S_{sc}\,W_c^{(0)\dagger}\,
    \xi^{(0)} + O(\lambda^5) \nonumber\\
   &+& \bigg\{ \bar\xi^{(0)}\,(i\Dslash_{c\perp}^{(0)}\,W_c^{(0)})\,
    S_{sc}^\dagger\,(1 + x_\perp\cdot D_{sc})\,\sigma
    + \bar\eta^{(0)}\,(i\Dslash_{c\perp}^{(0)}\,W_c^{(0)})\,
    S_{sc}^\dagger\,\theta \nonumber\\
   &&\mbox{}+ \bar\xi^{(0)}\,\frac{\nbslash}{2}\,(i n\cdot D_c^{(0)}\,
    W_c^{(0)})\,S_{sc}^\dagger\,\theta + \mbox{h.c.}
    + O(\lambda^{11/2}) \bigg\} \,,
\end{eqnarray}
where all soft-collinear fields are now evaluated at $x_-$. This result 
can be simplified further by introducing the gauge-invariant building 
blocks \cite{Hill:2002vw}
\begin{equation}
\begin{aligned}
   \X &= W_c^{(0)\dagger}\,\xi^{(0)}
    = S_{sc}^\dagger(x_-)\,W_c^\dagger\,\xi \,, \\
   \A_c^\mu &= W_c^{(0)\dagger}\,(iD_c^{(0)\mu}\,W_c^{(0)}) \\[-0.15cm]
   &= S_{sc}^\dagger(x_-)\,\Big[ W_c^\dagger\,(iD_c^\mu\,W_c)
    + \frac{\bar n^\mu}{2} \left( W_c^\dagger\,n\cdot A_{sc}(x_-)\,W_c
    - n\cdot A_{sc}(x_-) \right) \Big]\,S_{sc}(x_-) \,,
\end{aligned}
\end{equation}
which are invariant under both collinear and soft-collinear gauge 
transformations. Expressing the result in terms of these fields, and 
eliminating the field $\eta^{(0)}$ using its leading-order equation of 
motion
\begin{equation}
   \eta^{(0)} = - \frac{\nbslash}{2}\,\frac{1}{i\bar n\cdot D_c^{(0)}}\, 
   i\Dslash_{c\perp}^{(0)}\,\xi^{(0)} + \dots \,,
\end{equation}
we obtain the final result
\begin{eqnarray}\label{finalc}
   {\cal L}_{c+sc}^{\rm int}
   &=& \bar\X\,\frac{\nbslash}{2}\,S_{sc}^\dagger\,
    x_{\perp\mu} n_\nu\,g_s G_{sc}^{\mu\nu}\,S_{sc}\,\X + O(\lambda^5)
    \nonumber\\
   &+& \bigg\{ \bar\X\,\calAslash_{c\perp}\,S_{sc}^\dagger\,
    (1 + x_\perp\cdot D_{sc})\,\sigma
    - \bar\X\,i\calDslash_{c\perp}\,\frac{\nbslash}{2}\,
    \frac{1}{i\bar n\cdot\partial}\,\calAslash_{c\perp}\,
    S_{sc}^\dagger\,\theta + \bar\X\,\frac{\nbslash}{2}\,n\cdot\A_c\,
    S_{sc}^\dagger\,\theta \nonumber\\
   &&\mbox{}+ \mbox{h.c.} + O(\lambda^{11/2}) \bigg\} \,,
\end{eqnarray}
where $i\D_c^\mu=i\partial^\mu+\A_c^\mu$. Recall that the components of 
$x^\mu$ in this Lagrangian scale like $(1,\lambda^{-2},\lambda^{-1})$. 
Soft-collinear fields live at position $x_-$, while collinear fields live 
at position $x$. The small-component field $\sigma$ may be eliminated 
using (\ref{sigmasol}). Once again, the soft-collinear fields enter the 
result in combinations such as $S_{sc}^\dagger\,\theta$, which are 
explicitly gauge invariant.

In complete analogy to the $(s+sc)$ sector it follows that in terms of 
the redefined fields the interaction of two collinear quarks with a 
soft-collinear gluon (first line) is a subleading effect, for which we 
have computed the $O(\lambda^{1/2})$ contribution to the action. The 
interaction of a collinear quark and collinear gluon with a 
soft-collinear quark is also a subleading effect, for which we have 
computed the $O(\lambda^{1/2})$ and $O(\lambda)$ contributions to the 
action. In addition to the quark terms shown above there exist pure-glue 
interactions between soft-collinear and collinear gluons, which are again 
of subleading order in power counting. Their precise form will not be 
derived here.

\section{Induced soft-collinear interactions}
\label{sec:induced}

Above we have shown that there are no leading-order interactions between 
soft, collinear, and soft-collinear fields. The three interaction terms 
in (\ref{Lscet}) vanish at leading order in $\lambda$. This property of 
SCET$_{\rm II}$ is crucial to the idea of soft-collinear factorization, 
which is at the heart of QCD factorization theorems. In order to preserve 
a transparent power counting it is then convenient to define hadron 
states in the effective theory as eigenstates of one of the two 
leading-order Lagrangians ${\cal L}_s$ and ${\cal L}_c$. For instance, a 
``SCET pion'' would be a bound state of only collinear fields, and a 
``SCET $B$ meson'' would be a bound state of only soft fields. 

Since by definition these states do not contain any soft-collinear modes 
there is no need to include source terms for soft-collinear fields in the 
functional integral. ``Integrating out'' the soft-collinear fields from 
the path integral then gives rise to induced, highly non-local 
interactions between soft and collinear fields. The corresponding term in 
the action is of the form
\begin{equation}\label{SSC}
   S_{s+c}^{\rm induced} = i\int d^4x\,d^4y\,\mbox{T} \left\{
   {\cal L}_{c+sc}^{\rm int}(x)\,{\cal L}_{s+sc}^{\rm int}(y)\,
   e^{i\int d^4z\,{\cal L}_{sc}(z)} \right\}
\end{equation}
with all soft-collinear fields contracted. This result can be expressed
in terms of exact, gauge-invariant two-particle correlation functions of 
soft-collinear fields. 

Consider first the effective interaction between two soft and two 
collinear quarks arising from the exchange of a soft-collinear gluon. The 
result can be expressed in terms of the correlation function
\begin{eqnarray}
   &&\langle\Omega|\,\mbox{T} \left\{
    (S_{sc}^\dagger\,n_\nu\,g_s G_{sc}^{\mu\nu}\,S_{sc})_a(x_-)\,
    (W_{sc}^\dagger\,\bar n_\beta\,g_s G_{sc}^{\alpha\beta}\,
    W_{sc})_b(y_+)\,e^{i\int d^4z\,{\cal L}_{sc}(z)} \right\}
    |\Omega\rangle \nonumber\\
   &&\, = \delta_{ab}\,g_\perp^{\mu\alpha}\,\Delta_G(x_-\cdot y_+)
    + \dots \,,
\end{eqnarray}
where $a,b$ are color indices, and the dots represent terms that vanish 
when contracted with vectors in the transverse plane. SCET power counting 
implies that $\Delta_G(x_-\cdot y_+)\sim\lambda^6$. At tree level we 
find that $\Delta_G(x_-\cdot y_+)=-ig_s^2\,\delta^{(4)}(z)%
=-2ig_s^2\,\delta(x\cdot\bar n)\,\delta(y\cdot n)\,\delta^{(2)}(0)$, 
where $z=x_- - y_+$.~\footnote{In $d=4-2\epsilon$ dimensions, 
$\Delta_G=g_s^2\,\pi^{-2+\epsilon}\,\epsilon\,\Gamma(2-\epsilon)\,%
(2x_-\cdot y_+ + i0)^{-2+\epsilon}$.}
The resulting contribution to the induced interaction in (\ref{SSC}) is 
given by 
\begin{equation}\label{mess1}
   S_{s+c}^{\rm induced,(1)}
   = i\int d^4x\,d^4y\,x_\perp\cdot y_\perp\,\Delta_G(x_-\cdot y_+)\,
   (\bar\X\,\frac{\nbslash}{2}\,t_a\,\X)(x)\,
   (\bar\Q_s\,\frac{\nslash}{2}\,t_a\,\Q_s)(y) \,,
\end{equation}
which is of $O(\lambda)$ in power counting, as indicated by the 
superscript ``(1)''. 

Next we discuss the induced interactions obtained from the exchange of a
soft-collinear quark, as illustrated by the second diagram in 
Figure~\ref{fig:messenger}. The various correlation functions arising in 
the product of the relevant terms in (\ref{final}) and (\ref{finalc}) can 
all be related to the covariant expansion of the correlator
\begin{eqnarray}\label{eq:corr}
   &&\langle\Omega|\,\mbox{T} \left\{
   \Big[ S_{sc}^\dagger(x_-)\,(R_c^\dagger\,q_{sc})(x) \Big]_i\,
   \Big[ (\bar q_{sc}\,R_s)(y)\,W_{sc}(y_+) \Big]_j\,
   e^{i\int d^4z\,{\cal L}_{sc}(z)} \right\} |\Omega\rangle \nonumber\\
   &&\, = i\delta_{ij}\,\Big[ (\zslash_+ + \zslash_-)\,
    \Delta_1(z_+\cdot z_-,z_\perp^2)
    + \zslash_\perp \frac{\nbslash\nslash}{4}\,
    \Delta_2(z_+\cdot z_-,z_\perp^2) 
    + \zslash_\perp \frac{\nslash\nbslash}{4}\,
    \Delta_3(z_+\cdot z_-,z_\perp^2) \Big]
\end{eqnarray}
about the points $x=x_-$ and $y=y_+$. Here $i,j$ are color
indices. The form of the right-hand side of this equation holds to any
(finite) order in perturbation theory. Translational invariance
implies that the result is a function of $z=x-y$. Note, however, that
the presence of the light-cone vectors $n$ and $\bar n$ in the gauge
strings $S_{sc}$ and $W_{sc}$ breaks the rotational symmetry between
the longitudinal and transverse components of $z$. Nevertheless, the
expression is symmetric under the simultaneous interchange of
$n\leftrightarrow\bar n$ and $x\leftrightarrow y$ followed by
hermitean conjugation. Expanding the result (\ref{eq:corr}) to first
order in transverse displacements we find six non-zero
correlators, four of which are relevant to our analysis. They are
\begin{equation}
\begin{aligned}
   \langle\Omega|\,\mbox{T} \left\{
   (S_{sc}^\dagger\,\theta)_i(x_-)\,(\bar\theta\,W_{sc})_j(y_+)\,
   e^{i\int d^4z\,{\cal L}_{sc}(z)} \right\} |\Omega\rangle
   &= i\delta_{ij}\,\xslash_-\,\Delta_1(x_-\cdot y_+) \,, \\
   \langle\Omega|\,\mbox{T} \left\{
   (S_{sc}^\dagger\,\sigma)_i(x_-)\,(\bar\sigma\,W_{sc})_j(y_+)\,
   e^{i\int d^4z\,{\cal L}_{sc}(z)} \right\} |\Omega\rangle
   &= - i\delta_{ij}\,\yslash_+\,\Delta_1(x_-\cdot y_+) \,, \\
   \langle\Omega|\,\mbox{T} \left\{
   (S_{sc}^\dagger\,\sigma)_i(x_-)\,
   (\bar\theta\,(-i\overleftarrow{D}_{sc\perp}^\mu)\,W_{sc})_j(y_+)\,
   e^{i\int d^4z\,{\cal L}_{sc}(z)} \right\} |\Omega\rangle
   &= - \delta_{ij}\,\frac{\nbslash\nslash}{4}\,\gamma_\perp^\mu\,
    \Delta_2(x_-\cdot y_+) \,, \\
   \langle\Omega|\,\mbox{T} \left\{
   (S_{sc}^\dagger\,iD_{sc\perp}^\mu\,\sigma)_i(x_-)\,
   (\bar\theta\,W_{sc})_j(y_+)\,
   e^{i\int d^4z\,{\cal L}_{sc}(z)} \right\} |\Omega\rangle
   &= - \delta_{ij}\,\frac{\nbslash\nslash}{4}\,\gamma_\perp^\mu\,
    \Delta_2(x_-\cdot y_+) \,,
\end{aligned}
\end{equation}
where $\Delta_n(x_-\cdot y_+)\equiv\Delta_n(-x_-\cdot y_+,0)$. ``Mixed'' 
correlators without an extra transverse derivative such as $\langle\,%
(S_{sc}^\dagger\,\sigma)(x_-)\,(\bar\theta\,W_{sc})(y_+)\,\rangle$ vanish 
due to the fact that, according to (\ref{Lsc}) and (\ref{sigmasol}), they 
involve an odd number of $\gamma_\perp$ matrices but no transverse 
Lorentz index. SCET power counting implies that the functions 
$\Delta_n(x_-\cdot y_+)\sim\lambda^6$. In fact, at tree level we obtain 
$\Delta_n(x_-\cdot y_+)=1/[8\pi^2(x_-\cdot y_+)^2]$ for 
$n=1,2,3$.~\footnote{In $d=4-2\epsilon$ dimensions, the expression is
$\frac12\,\pi^{-2+\epsilon}\,\Gamma(2-\epsilon)\,%
(2x_-\cdot y_+ + i0)^{-2+\epsilon}$.}
(Recall that $x_-\sim\lambda^{-2}$ and $y_+\sim\lambda^{-1}$.) It is now 
straightforward to evaluate the corresponding terms in the action 
(\ref{SSC}). We find
\begin{eqnarray}
   S_{s+c}^{\rm induced,(3/2)}
   &=& \int d^4x\,d^4y\,\Delta_1(x_-\cdot y_+)\,\bigg\{
    \bar\X\,\calAslash_{c\perp}\,\yslash_+ 
    \left( \frac{\nslash}{2}\,\bar n\cdot\A_s + \calAslash_{s\perp}
    \right) \Q_s \nonumber\\
   &&\quad\mbox{}- \bar\X \left( \frac{\nbslash}{2}\,n\cdot\A_c
    - i\calDslash_{c\perp}\,\frac{1}{i\bar n\cdot\partial}\,
    \frac{\nbslash}{2}\,\calAslash_{c\perp} \right)
    \xslash_-\,\calAslash_{s\perp}\,\Q_s \bigg\} \nonumber\\
   &-& \int d^4x\,d^4y\,\Delta_2(x_-\cdot y_+)\,
    \bar\X\,\calAslash_{c\perp}\,(\xslash_\perp-\yslash_\perp)\,
    \calAslash_{s\perp}\,\Q_s + \mbox{h.c.} \,,
\end{eqnarray}
where it is understood that all collinear fields live at $x$, while all 
soft fields live at $y$. Power counting shows that this induced long-range
soft-collinear interaction is of $O(\lambda^{3/2})$, as indicated by the 
superscript. We stress that the superficially leading term of 
$O(\lambda)$ vanishes due to rotational invariance in the transverse 
plane. 

At the same order in power counting there appear terms in the interaction 
Lagrangian ${\cal L}_{s+c}^{\rm int}$ in (\ref{Lscet}) from integrating 
out hard-collinear modes. At tree level, hard-collinear gluon exchange 
induces the interaction
\begin{equation}\label{Lsc1}
   {\cal L}_{s+c}^{\rm int,(1)} = -4\pi\alpha_s
   \left[ \bar\X(x_+ + x_\perp)\,
   \frac{\gamma_\mu\,t_a}{in\cdot\partial}\,\Q_s(x_- + x_\perp) \right]
   \left[ \bar\Q_s(x_- + x_\perp)\,
   \frac{\gamma^\mu\,t_a}{i\bar n\cdot\partial}\,\X(x_++x_\perp) \right] ,
\end{equation}
which should be understood as a matching contribution to the effective
Lagrangian at the hard-collinear scale $\mu^2\sim E\Lambda$. This 
operator (in Fierz-transformed form) has been derived previously in a
discussion of color-suppressed hadronic $B$ decays in 
\cite{Mantry:2003uz}. Power counting shows that it scales like 
$\lambda^4$, which when combined with the measure $d^4x\sim\lambda^{-3}$ 
indeed gives rise to an $O(\lambda)$ interaction term in the action. Note 
that in terms containing both soft and collinear fields, the soft fields 
must be multipole expanded about $x_+=0$, while the collinear fields must 
be expanded about $x_-=0$. This point was not emphasized in 
\cite{Hill:2002vw}. The operator obtained from the exchange of a 
hard-collinear quark shown in Figure \ref{fig:messenger} has the form
\begin{eqnarray}\label{Lsc2}
   {\cal L}_{s+c}^{\rm int,(3/2)}
   &=& - \bar\X\,\frac{1}{i\bar n\cdot\partial}\,\calAslash_{s\perp}\,
    \frac{\nbslash}{2}\,\calAslash_{c\perp}\,\Q_s
    - \bar\X\,\frac{1}{i\bar n\cdot\partial}\,\calAslash_{s\perp}\,
    (i\delslash_\perp + \calAslash_{s\perp} + \calAslash_{c\perp})\,
    \calAslash_{c\perp}\,\frac{1}{in\cdot\partial}\,\Q_s \nonumber \\
   &&\mbox{}\!- \bar\X\,(i\delslash_\perp + \calAslash_{c\perp})\,
    \frac{1}{i\bar n\cdot\partial}\,\calAslash_{s\perp}\,
    \calAslash_{c\perp}\,\frac{1}{i n\cdot\partial}\,\Q_s
    + \mbox{h.c.} \,,
\end{eqnarray}
where, as in (\ref{Lsc1}), all collinear fields are to be evaluated at 
the point $x_+ + x_\perp$ and the soft fields at $x_- + x_\perp$. 

\begin{figure}
\begin{center}
\psfrag{l2}[b]{\scriptsize $~m_b v+l_2$}
\psfrag{p1}[b]{\scriptsize $p_2$}
\psfrag{p2}[B]{\scriptsize $p_1$}
\psfrag{l1}[Bl]{\scriptsize $~l_1$}
\psfrag{B}[B]{\small\phantom{,}$B$}
\psfrag{M}[B]{\small\phantom{i} $M$}
\begin{tabular}{ccccc}
\raisebox{-0.025\textwidth}{%
\includegraphics[height=0.18\textwidth]{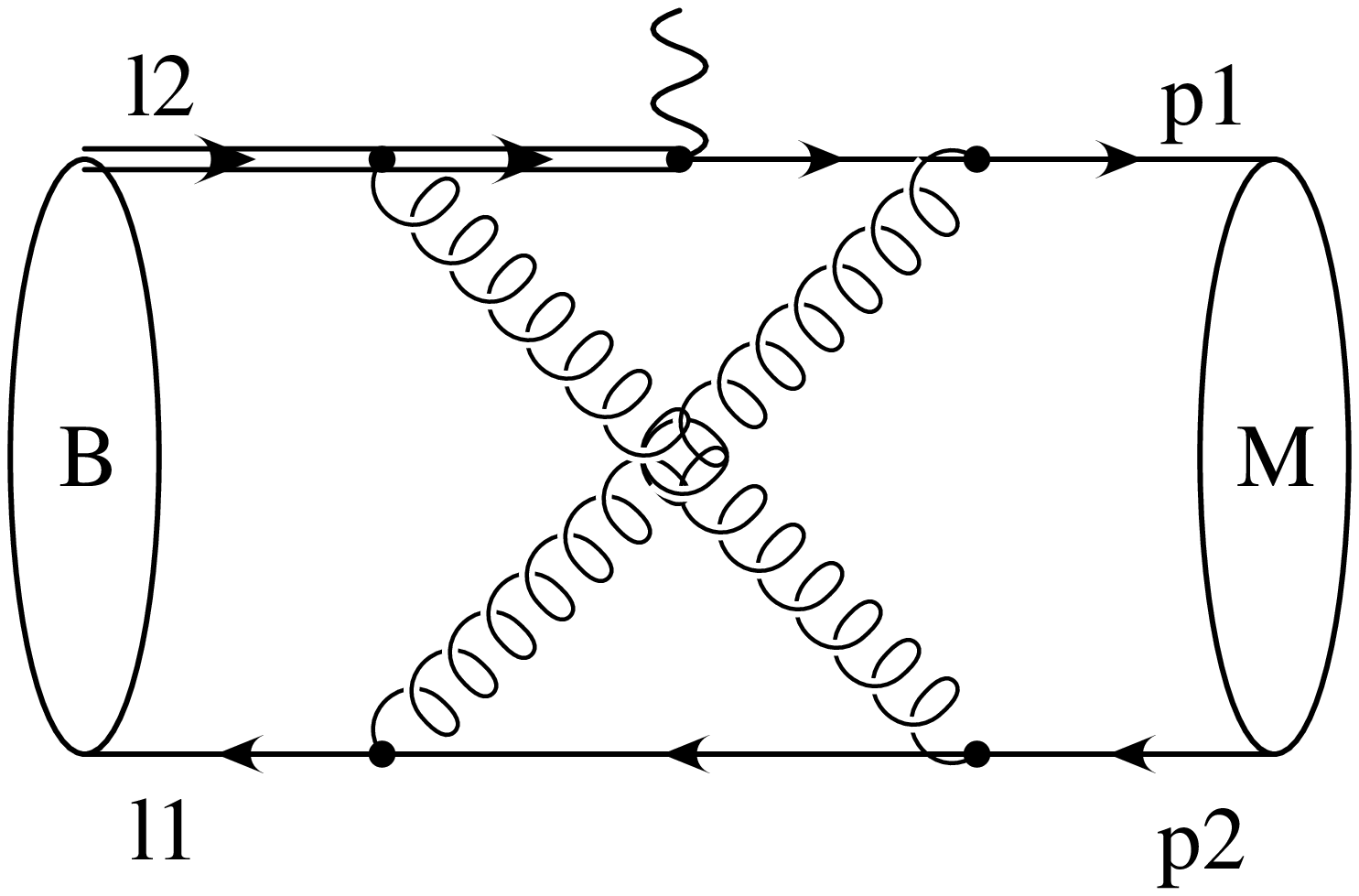}} 
& \raisebox{0.06\textwidth}{$\rightarrow$}
& \includegraphics[height=0.13\textwidth]{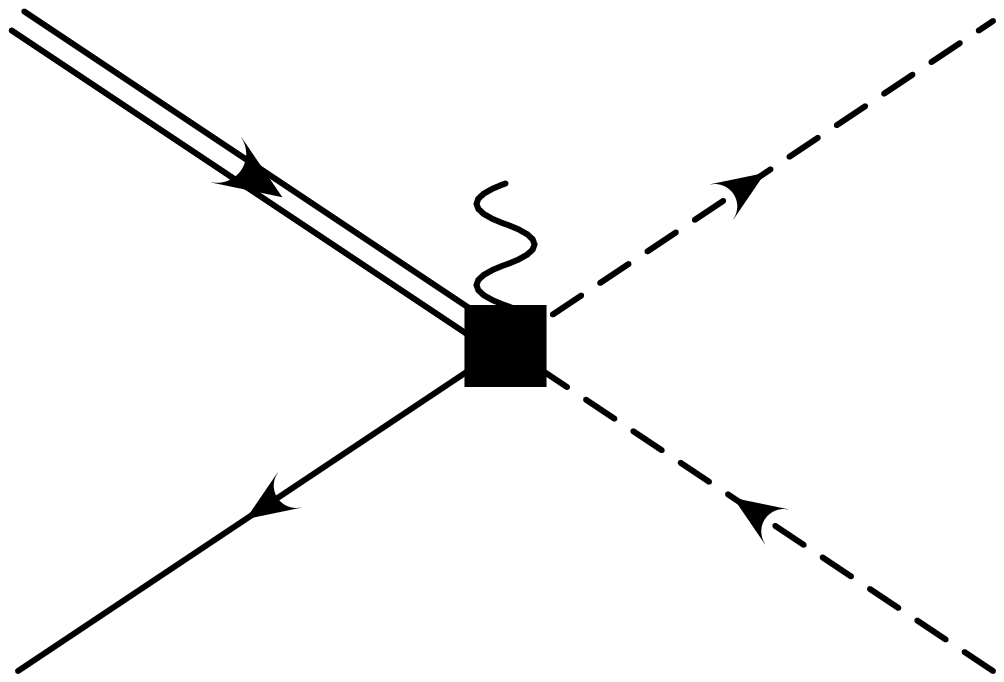} 
& \raisebox{0.06\textwidth}{$+\,\dots\,+$}
& \includegraphics[height=0.16\textwidth]{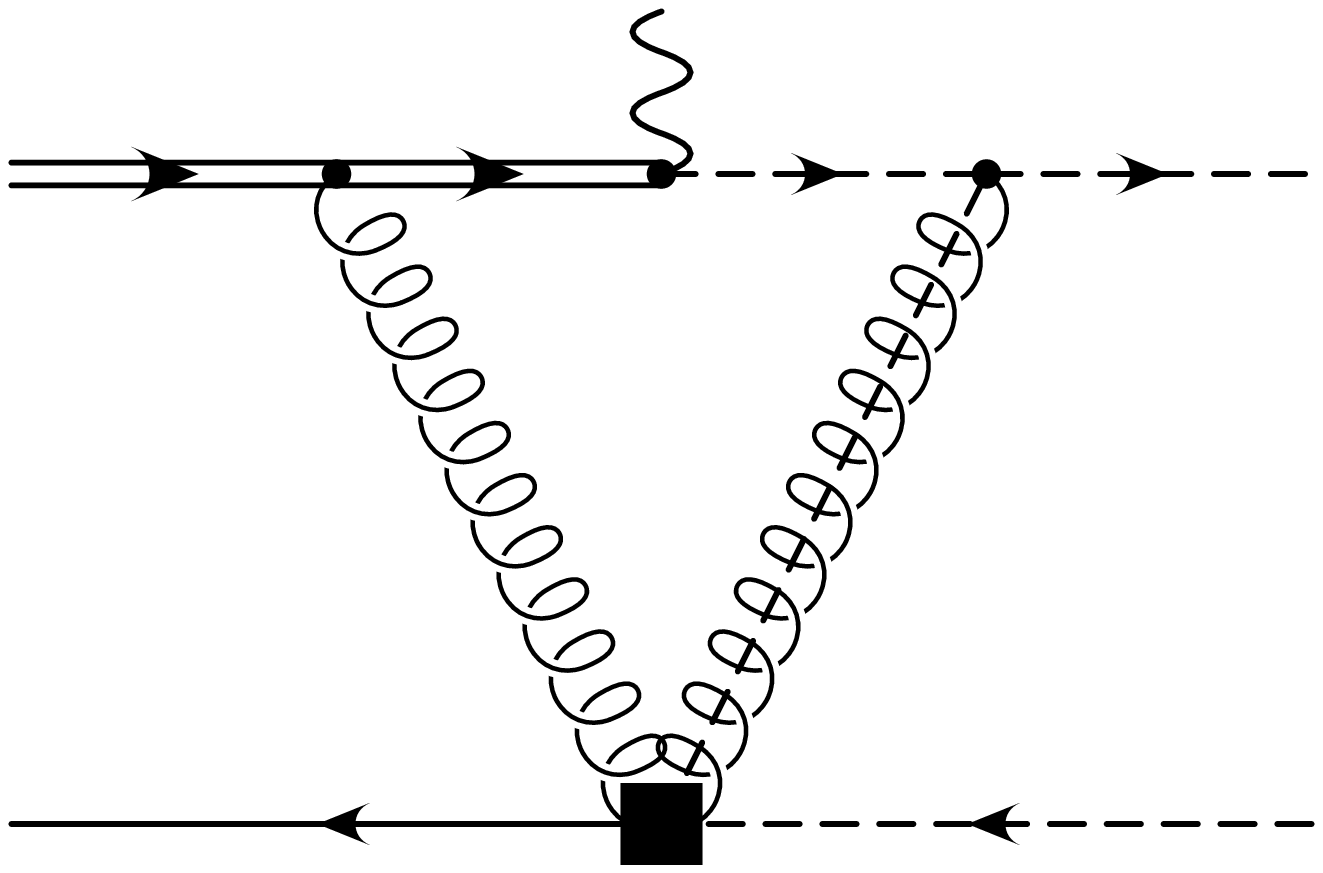}
\end{tabular}
\end{center}
\vspace{-0.2cm} 
\centerline{\parbox{14cm}{\caption{\label{fig:hardcoll}
A QCD diagram contributing to the decay of a $B$ meson to an energetic
light meson $M$, and its representation in the effective theory. Solid
lines carry soft, dashed lines collinear momentum. External soft momenta 
are incoming and collinear ones outgoing. The last diagram, which 
involves the interaction (\ref{Lsc2}), vanishes.}}}
\end{figure}

Translated to momentum space, this particular position dependence of the 
fields enforces that the minus component of the total collinear momentum 
flowing into the vertex exits again through collinear lines. An analogous 
statement holds for the plus component of the soft momenta. The 
momentum-conservation $\delta$-functions associated with the above 
vertices therefore reads 
$\delta(\bar n\cdot P)\,\delta(n\cdot L)\,\delta^{(2)}(P_\perp+L_\perp)$, 
where $P$ is the sum of all collinear and $L$ the sum of all soft momenta
(ingoing minus outgoing) connected to the vertex. As classical scattering 
processes, these interactions are rather exceptional: the 
operator (\ref{Lsc1}), for example, contributes only to the forward 
scattering of a collinear quark and a soft quark. When inserted into loop 
diagrams, such exceptional momentum configurations in general do not give 
rise to non-zero contributions. To see this, let us analyze the 
one-loop QCD diagram shown in Figure~\ref{fig:hardcoll} using the strategy of 
regions. One finds that the graph receives a contribution from the region 
where all propagators are hard or hard-collinear. In the effective theory 
this contribution is represented by the first diagram on the right-hand 
side. There are also contributions from regions where two propagators of 
the loop are soft (or collinear) and all others hard or hard-collinear. 
These correspond to effective-theory diagrams (not shown) where a 
collinear or soft gluon is emitted from the weak vertex and absorbed by 
one of the quarks. However, no contribution arises that would correspond 
to the last diagram in Figure~\ref{fig:hardcoll}, which involves the 
interaction (\ref{Lsc2}) denoted by the black square. First, it is 
impossible to assign a loop momentum and external momenta in the 
full-theory diagram such that the exceptional momentum configuration 
corresponding to the last graph (where the gluon connected to the heavy 
quark is soft and that connected to the outgoing light quark is 
collinear) would be enforced. This configuration would require the loop
momentum to scale like $k^\mu\sim(\lambda^2,\lambda,\lambda)$, in which 
case the denominators of the QCD loop integral can be simplified as
\begin{equation}
\begin{aligned}
   &\frac{1}{2l_{1+}\cdot(k_- + l_{1-}) + (k_\perp+l_{1\perp})^2 + i0}\,
    \,\frac{1}{v\cdot(k_- + k_\perp + l) + i0} \\
   \times~
   & \frac{1}{2p_{1-}\cdot(k_+ + p_{1+}) + (k_\perp+p_{1\perp})^2 + i0}\,
    \,\frac{1}{2p_{-}\cdot(k_+ + p_+) + (k_\perp+p_{\perp})^2 + i0} \,,
\end{aligned}
\end{equation}
where $l=l_1+l_2$ and $p=p_1+p_2$. These denominator structures precisely
correspond to the multipole-expanded form of the operator (\ref{Lsc2}).
Note that the plus component of the loop momentum does not propagate into
soft lines, while its minus component does not propagate into collinear
lines. As a result, the integration over $k_-$ vanishes if the components 
$n\cdot l_i$ of the external soft momenta all have the same sign, because 
then the contour can be deformed away from the poles. Similarly, the 
integration over $k_+$ vanishes if the components $\bar n\cdot p_i$ of 
the external collinear momenta all have the same sign. The same 
conclusion is reached upon analyzing a general loop diagram involving 
(\ref{Lsc1}) or (\ref{Lsc2}): non-zero contributions from these 
interactions can only occur in processes which involve soft and collinear 
particles in both the initial and final state, and these interactions 
therefore do not contribute in decays of $B$ mesons. This is a consequence 
of the Coleman--Norton theorem \cite{Coleman}, which states that on-shell 
singularities in Feynman diagrams correspond to classically allowed 
scattering processes. The above conclusion also applies to the analysis 
in \cite{Mantry:2003uz}. The non-perturbative soft-collinear rescattering 
mechanism introduced there should not be described in terms of 
(\ref{Lsc1}), but rather through the long-distance messenger exchange 
interaction (\ref{mess1}).

As a final note, let us mention that the same arguments ensure that there 
are no induced self-interactions of the type (\ref{SSC}) among only soft 
or only collinear fields, i.e., the effective Lagrangians ${\cal L}_s$ 
and ${\cal L}_c$ are not altered by integrating out the soft-collinear 
modes. The corresponding correlation functions $\langle\,%
(S_{sc}^\dagger\,\theta)(x_-)\,(\bar\theta\,S_{sc})(y_-)\,\rangle$ etc.\ 
vanish, since the contour can be closed avoiding the poles in the complex 
plane. This ensures that soft-collinear fields can only be exchanged in 
graphs involving both soft and collinear fields.

\section{Conclusions}

We have argued that the version of soft-collinear effective theory 
appropriate for the discussion of exclusive $B$ decays such as 
$B\to\pi\,l\,\nu$, $B\to K^*\gamma$, and $B\to\pi\pi$ should include a 
new mode in addition to soft and collinear fields, which can interact 
with both soft and collinear particles without taking them far off-shell. 
The relevance of these messenger fields has been demonstrated using the 
example of a triangle diagram with one soft and one collinear external 
momentum. We have then developed the formalism incorporating the 
corresponding quark and gluon fields into the soft-collinear effective 
Lagrangian. 

In the strong-interaction sector of the effective theory the 
leading-order interactions of soft-collinear fields with soft or 
collinear fields can be removed using field redefinitions. The remaining
interactions are power suppressed. We have explicitly worked out the 
leading terms involving quark fields. We stress that this decoupling in 
the QCD sector of the theory does not necessarily imply that 
soft-collinear modes can be ignored at leading order in power counting. 
This would only be true in cases where the decoupling transformation 
leaves external operators such as weak-interaction currents invariant. 
When the messenger fields are ``integrated out'' in the functional 
integral they give rise to highly non-local induced interactions between 
soft and collinear fields, which appear at the same order in power 
counting as short-distance interactions among these fields induced by 
the exchange of hard-collinear modes. However, only the long-distance 
interactions induced by soft-collinear exchange graphs are kinematically 
allowed in $B$ decays, where no collinear particles are present in the 
initial state.

A puzzling aspect of our analysis is the finding of a relevant momentum 
region corresponding to virtualities that are parametrically below the 
QCD scale $\Lambda_{\rm QCD}$. However, our discussion in 
Section~\ref{sec:Sudakov} illustrates that the same phenomenon arises in 
the case of the Sudakov form factor, if the external massless particles 
have off-shell momenta scaling like $p_1^\mu\sim(\Lambda^2/E,E,\Lambda)$
and $p_2^\mu\sim(E,\Lambda^2/E,\Lambda)$ with 
$\Lambda\sim\Lambda_{\rm QCD}$. Then perturbatively the ultra-soft region
of loop momenta $k^\mu\sim\Lambda^2/E$ gives rise to leading-order 
contributions and acts as a ``messenger'' between the two collinear 
particles, just like the soft-collinear modes connecting soft and 
collinear particles in our case.

The fact that modes with $p^2\ll\Lambda^2$ represent very long-range 
fluctuations compared to the scale of hadronic systems is not necessarily
an argument against their existence. It is to some extent a 
consequence of dimensional regularization and analyticity that the 
relevant momentum region is $k^\mu\sim E(\lambda^2,\lambda,\lambda^{3/2})$
and not $k^\mu\sim E(\lambda^2,\lambda,\lambda)$, in which case the 
virtuality would be of order $\Lambda^2$. Ultimately, it is not the 
virtuality that is important but the fact that the plus and minus 
components of soft-collinear momenta are commensurate with certain 
components of collinear or soft momenta. As an analogy, let us recall the 
case of the hydrogen Lamb shift, in which ``ultra-soft'' modes with time 
and space components of momentum scaling like $(mv^2,mv^2)$ are 
important. These modes have virtuality $p^2\ll(mv)^2$, where $mv$ is the 
typical momentum of the hydrogen atom. Also in this case it is not the 
off-shellness that matters, but rather the fact that one of the 
components (the energy) is comparable to that of the bound-state system.

Finally, let us mention that the introduction of a mass term 
$m\sim\lambda$ in the propagator labeled with $k$ in 
Figure~\ref{fig:triangle} would change the analysis of regions. In this 
case there would be no contribution from the soft-collinear region, since 
the loop momentum $k^2\sim\lambda^3$ in the soft-collinear propagator 
$1/(k^2-m^2)$ could be Taylor-expanded and no pinch singularities would 
occur. However, this ``simplification'' comes at a price: after the 
introduction of the mass term, the loop integrals occurring in the soft 
and collinear region are no longer regulated dimensionally. In order to 
obtain the expansion of the massive integral using the strategy of 
regions, additional analytic regulators have to be introduced on the soft 
and collinear propagators \cite{Smirnov:1998vk}. The procedure is 
straightforward as long as one is only interested in the expansion of a 
given loop integral, but would complicate the construction of the 
effective theory since the analytic regulators destroy longitudinal boost 
and gauge invariance. We believe it is fair to say that to date it is not 
known how to consistently incorporate masses of $O(\Lambda)$ into 
soft-collinear effective theory.

A related question one might worry about is whether the scaling laws
for the propagator functions $\Delta_n(x_-\cdot y_+)$ describing the
soft-collinear exchange, which follow from the power-counting rules of
the effective theory, could be invalidated by some non-perturbative
effects not seen at the level of Feynman diagrams, such as a generation 
of mass terms in exact propagators. In our opinion such effects would not 
only upset power counting but threaten the usefulness of the effective 
theory as a whole. For now we take the conservative point of view that in
order to be useful the effective theory should at least reproduce 
correctly the analytic properties of perturbation theory. As we have 
shown, the inclusion of soft-collinear modes is then an indispensable 
part in the construction of the effective Lagrangian.

\vspace{0.5cm}
{\bf Acknowledgments:}
One of us (M.N.) is grateful to the Aspen Center of Physics, where part 
of this work was carried out. We would like to thank B.~Lange, C.~Bauer 
and I.~Stewart for useful discussions. The work of T.B.\ and R.J.H.\ is 
supported by the Department of Energy under Grant DE-AC03-76SF00515. The 
research of M.N.\ is supported by the National Science Foundation under 
Grant PHY-0098631.

\end{document}